\documentclass{aastex62}
\newcommand\asat{{\it AstroSat}}
\newcommand\xte{{\it RXTE}}
\newcommand\swift{{\it Swift}}

\newcommand\chan{{\it Chandra}}
\newcommand\maxi{{\it MAXI}}
\newcommand\nus{{\it NuSTAR}}
\newcommand\suzaku{{\it Suzaku}}
\newcommand\xmm{{\it XMM-Newton}}
\newcommand{\lsim}{\raisebox{-0.3ex}{\mbox{$\stackrel{<}{_\sim} \,$}}}

\usepackage{epsfig}
\usepackage{amsmath}
\usepackage{amssymb}
\usepackage{natbib}
\usepackage{graphicx}
\usepackage{color}

\makeatletter
\newcommand\footnoteref[1]{\protected@xdef\@thefnmark{\ref{#1}}\@footnotemark}
\makeatother

\shorttitle{AstroSat and Chandra observations of 4U 1630--47}
\shortauthors{Pahari et al.}

\begin{document}

\title{\asat{} and \chan{} view of the high soft state of 4U 1630--47 (4U 1630--472): evidence of the disk wind and a rapidly spinning black hole}

\correspondingauthor{Mayukh Pahari}
\email{M.Pahari@soton.ac.uk}

\author{Mayukh Pahari}
\affiliation{School of Physics and Astronomy, University of Southampton, Southampton, SO17 1BJ, UK}
\affiliation{Royal Society-SERB Newton International Fellow}
\affiliation{Department of Astronomy and Astrophysics, Tata Institute of Fundamental Research, Mumbai 400005, India}

\author{Sudip Bhattacharyya}
\affiliation{Department of Astronomy and Astrophysics, Tata Institute of Fundamental Research, Mumbai 400005, India}

\author{A R Rao}
\affiliation{Department of Astronomy and Astrophysics, Tata Institute of Fundamental Research, Mumbai 400005, India}

\author{Dipankar Bhattacharya}
\affiliation{Inter-University Centre for Astronomy and Astrophysics, Ganeshkhind, Pune 411 007, India}

\author{Santosh V Vadawale}
\affiliation{Physical Research Laboratory, 380009, Ahmedabad, Gujarat, India}

\author{Gulab C Dewangan}
\affiliation{Inter-University Centre for Astronomy and Astrophysics, Ganeshkhind, Pune 411 007, India}

\author{I M McHardy}
\affiliation{School of Physics and Astronomy, University of Southampton, Southampton, SO17 1BJ, UK}

\author{Poshak Gandhi}
\affiliation{School of Physics and Astronomy, University of Southampton, Southampton, SO17 1BJ, UK}

\author{St\'ephane Corbel}
\affiliation{University Paris Diderot and CEA Saclay and Observatoire de Paris, France}

\author{Norbert S Schulz}
\affiliation{Kavli Institute for Astrophysics and Space Research, Massachusetts Institute of Technology, Cambridge, MA 02139, USA}

\author{Diego Altamirano}
\affiliation{School of Physics and Astronomy, University of Southampton, Southampton, SO17 1BJ, UK}

\begin{abstract}
We present the X-ray spectral and timing analysis of the transient black hole X-ray binary 4U 1630--47, observed with the \asat{}, \chan{} and \maxi{} space missions during its soft X-ray outburst in 2016. The outburst, from the rising phase until the peak, is neither detected in hard X-rays (15-50 keV) by the \swift{}/BAT  nor in radio. Such non-detection along with the source behavior in the hardness-intensity and color-color diagrams obtained using \maxi{} data confirm that both \chan{} and \asat{} observations were performed during the high soft spectral state. The High Energy Grating (HEG) spectrum from the \chan{} high-energy transmission grating spectrometer (HETGS) shows two strong, moderately blueshifted absorption lines at 6.705$_{-0.002}^{+0.002}$ keV and 6.974$_{-0.003}^{+0.004}$ keV, which are produced by Fe XXV and Fe XXVI in a low-velocity ionized disk wind. The corresponding outflow velocity is determined to be 366$\pm$56 km/s. Separate spectral fits of \chan{}/HEG, \asat{}/SXT+LAXPC and \chan{}/HEG+\asat{}/SXT+LAXPC data show that the broadband continuum can be well described with a relativistic disk-blackbody model, with the disk flux fraction of $\sim 0.97$. Based on the best-fit continuum spectral modeling of \chan{}, \asat{} and \chan{}+\asat{} joint spectra and using the Markov Chain Monte Carlo simulations, we constrain the spectral hardening factor at 1.56$^{+0.14}_{-0.06}$ and the dimensionless black hole spin parameter at 0.92 $\pm$ 0.04 within the 99.7\% confidence interval. Our conclusion of a rapidly-spinning black hole in 4U 1630--47 using the continuum spectrum method is in agreement with a previous finding applying the reflection spectral fitting method.
\end{abstract}
  

\section{Introduction}\label{Introduction}

A transient black hole X-ray binary (BHXB) traces several X-ray spectral states during its outburst. Among these states, physical processes associated with the high-intensity, soft, thermal emission-dominated state or the high soft (HS) state are possibly the best understood, due to the relatively simple, non-degenerate spectral modeling. 
The HS state emission is dominated by a blackbody emission from a geometrically thin and optically thick accretion disk, with the ratio of the disk flux to the total flux greater than 70\%$-$80\% \citep{re06}. 
In particular, the Shakura-Sunyaev prescription \citep{sh73} predicts that the emitting disk may extend down to the innermost stable circular orbit (ISCO) around the black hole.
In addition to the disk-blackbody emission, a low luminosity, non-thermal, hard power-law tail with a variable photon index $>$ 1.75 \citep[may reach $\sim 5$; ][]{ti10,mo09,re06} has also been observed in the HS state. Such a tail is often explained by the presence of `patchy,' magnetized, active coronal blobs, which are largely independent of underlying accretion and Comptonization details around the thin disk \citep{Ha94}.

Using a large number of HS state spectra obtained from different outbursts in X-ray binaries using different satellites, it has been found that the observed disk luminosity and the observed disk temperature follow an L $\propto$ T$^4$ relationship, and the apparent inner disk radius remains constant in spite of a variable disk luminosity \citep{St10,Mu99}. Therefore, it is reasonable to assume that such a stable inner disk radius is the ISCO radius ($r_{\rm ISCO}$). For the Kerr spacetime, and assuming a corotating disk, the black hole dimensionless spin parameter $a_*$ ($= cJ/GM^2$) can be uniquely estimated from a measured value of $r_{\rm ISCO}/M$, where $M$ and $J$ are the black hole mass and total angular momentum respectively \citep[e.g., ][]{Mc06}. Therefore, fitting of the thermal component of the continuum spectrum has been used to estimate $r_{\rm ISCO}/M$, and hence $a_*$, of several accreting black holes \citep[e.g., ][]{Sha08,FragosMcClintock2015}. The continuum spectrum fitting method is often used when the source inclination angle ($i$), distance ($D$) and $M$ are independently known. The method is particularly promising, if almost the entire X-ray emission is from the
disk, which significantly reduces systematic uncertainties.
Note that $r_{\rm ISCO}/M$, and hence $a_*$, can also be estimated by an alternative method, viz., the fitting of the energy and shape of a broad relativistic spectral line. Such a fluorescent line is believed to originate from the reflection of hard X-rays off the inner disk, and the line is shaped by the Doppler effect, special relativistic beaming, gravitational redshift, etc.  \citep[e.g., ][]{Mi07}.

\begin{figure}
\begin{center}
\includegraphics[scale=0.5]{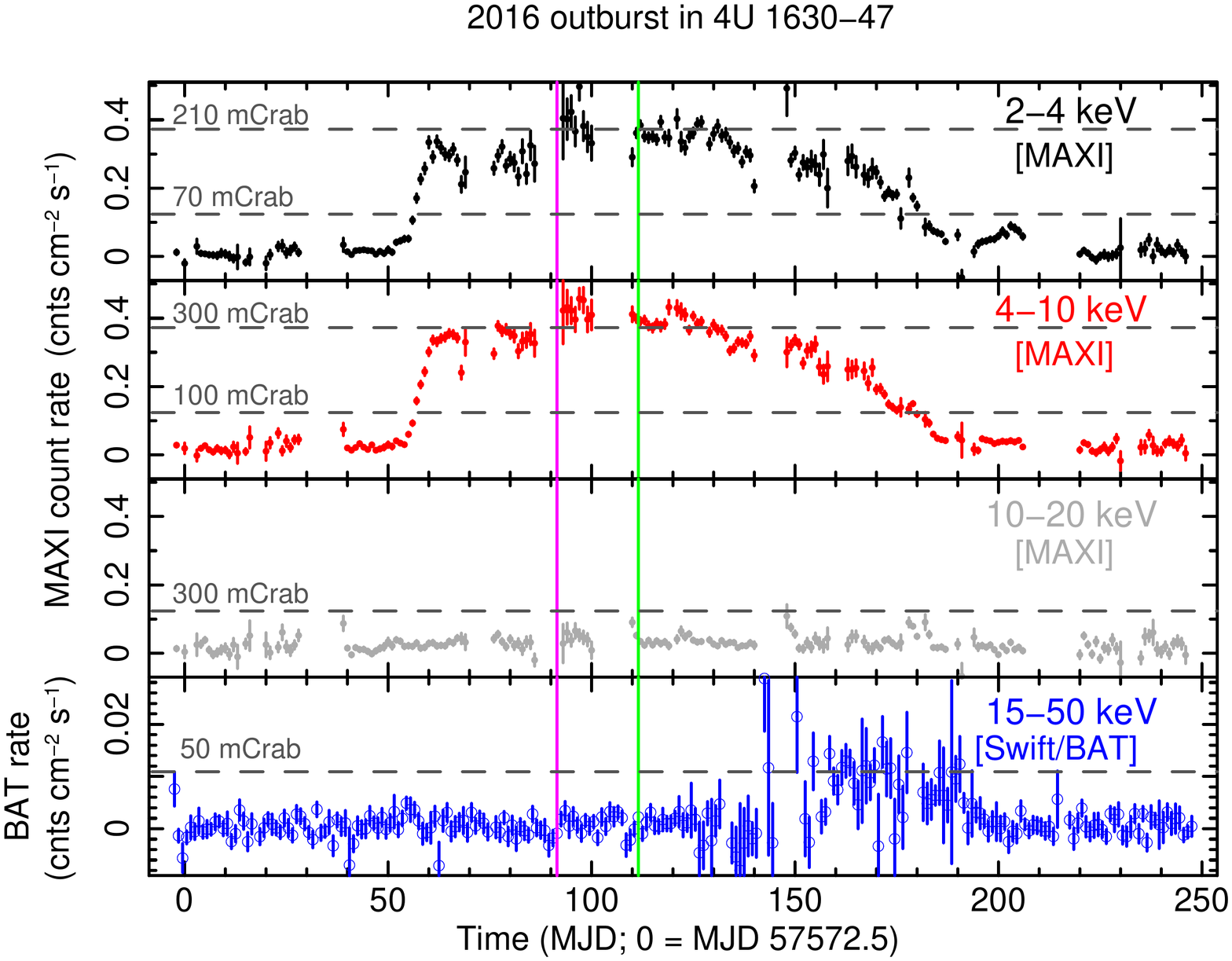}
\caption{2016 X-ray outburst of 4U 1630--47 as observed by \maxi{} and \swift{}/BAT. Top, upper-middle and lower-middle panels show 2-4 keV, 4-10 keV and 10-20 keV one-day averaged lightcurve from \maxi{}. Interestingly the outburst in 10-20 keV is close to the non-detection level of \maxi{}. Bottom panel shows 1 day averaged, 15-50 keV \swift{}/BAT lightcurve of the same outburst. 4U 1630--47 is marginally detected by \swift{}/BAT only during the decay phase of the outburst. The vertical left and right lines show the times of the \asat{} and \chan{} observations analyzed in this work (see Section \ref{state}).} 
\label{maxilight}
\end{center}
\end{figure}

The black hole HS state is also ideal for observing signatures of wind from the accretion disk \citep[e.g., ][]{Ne09}. Such features are usually narrow, blueshifted spectral lines which can be observed with \chan{} gratings. These lines can provide information about the wind velocity, ionization state and can generally be useful for understanding the matter inflow-outflow mechanism.

In this paper, we focus on the transient BHXB 4U 1630--47 \citep{jo76, pa95},
which goes into outbursts with a typical interval of 600--690 days \citep{ab05,to14}.
No dynamical mass measurement was possible for 4U 1630--47, because the optical
counterpart has not been identified due to the high extinction in the 
direction of the source near the Galactic plane. Therefore,
the identification of the source as a black hole \citep{pa86} was based on the 
similarity of its spectral and timing properties with those of BHXBs with measured 
black-hole masses \citep[e.g.,][]{ba96,ab05}.
Nevertheless, using the correlation between photon index, low-frequency quasi-periodic oscillations (QPOs) and mass accretion rate, \citet{se14} estimated the compact object mass and inner disk inclination angle of 10 $\pm$ 0.1 M$_\odot$ and $\lsim$ 70$^\circ$ respectively.
Note that the observations of flux dips, but a lack of full eclipses, imply a high (but not edge-on) inclination angle \citep[$\sim 70^\circ$; ][]{ki14,to98,Kuulkersetal1998} for 4U 1630--47, which is in agreement with the finding of \citet{se14}.

While the distance of 4U 1630--47 is not known with certainty, the presence of heavy absorption \citep[hydrogen column density $N_{\rm H}=5-12 \times 10^{22}$ cm$^{-2}$; ][]{to98}, faint optical counterpart ($>$ 20 magnitude), probably due to the high extinction in the Galactic plane \citep[monochromatic extinction at 5495 \AA ~is $>$ 9 at the Galactic latitude of $<$5$^\circ$;][] {sa14} and reddening \citep{se14} support a large distance ($>$ 8 kpc) of the source. Moreover, in the direction of 4U 1630--47, the presence of a Giant Molecular Cloud at a measured distance of 11 kpc \citep{au01} behind the source is consistent with the assumed distance to the source and provides an upper limit to the distance of the source.

The black hole spin $a_*$ value for 4U 1630--47, to the best of our knowledge, has so far not been estimated using the continuum spectrum method mentioned above.
However, \citet{ki14} reported, for the first time from this source, a broad
relativistic spectral iron emission line using \nus{} data. 
During this \nus{} observation, 4U 1630--47 was in an intermediate state,
showing the signature of an accretion disk, a hard Comptonized tail and
reflection  features from the inner disk. By fitting the reflection spectrum, \citet{ki14}
estimated $a_* = 0.985^{+0.005}_{-0.014}$ ($1\sigma$ statistical errors), and
$i \approx 64^\circ\pm2^\circ$. 

\suzaku{} spectra of 4U 1630--47 showed H-like and He-like Fe absorption lines at $\sim$6.96 keV and $\sim$6.7 keV \citep{ku07}.
These lines were thought to originate from the strongly ionized material in a disk wind.
However, using the same \suzaku{} spectra, \citet{ro14} proposed the accretion disk atmosphere as an alternative origin of the absorption lines.
A signature of thermally/radiatively driven disk wind with its ionization having a positive correlation with the source luminosity was noted by \citet{di14} using \xmm{} spectra of 4U 1630--47 during the 2012-2013 outburst. They found that the absorption features disappeared and an emission line appeared at very high luminosity, which implies a significant change in the degree of ionization of wind with increasing luminosity. 
Later, using a simpler model for the same \xmm{} spectra, \citet{wa16} showed the presence of a variable element abundance and a strongly-ionized absorber near the black hole. 
Using \chan{}/HETGS grating spectra of 4U 1630--47, \citet{ne13,ne14} found robust and strong absorption line caused by the outflowing wind launched during the outburst phase of 4U 1630--47. \citet{ne14} showed that the detection of ultra-relativistic wind from 4U 1630--47 \citep{di13} is ambiguous and the radio emission from this system may be unrelated to the X-ray emission lines. \citet{mi15} found a disk wind that requires magnetic launching from  two absorption zones at different radii having velocities of $\sim \rm{270~km~s^{-1}}$ and $\sim \rm{2100~km~s^{-1}}$. 

With this background, here we study the 2016 X-ray outburst of 4U 1630--47 in its high soft state. 
We report the results from the first \asat{} observation of this source. Besides, we analyze \chan{} data of the same outburst. 
These provide us with a rare opportunity to fit high spectral resolution grating spectra and the $\sim 0.3-80$~keV broadband spectra in the same state of the same outburst.
Using this \chan{} grating and \asat{} broadband spectra, and with the continuum spectrum method, we confirm the previously inferred high $a_*$ of the black hole.
We also report strong and blueshifted Fe XXV and Fe XXVI absorption lines in the \chan{} grating spectra,
implying a low velocity, ionized disk wind in 4U 1630--47 during our observation.

\begin{figure}
\begin{center}
\includegraphics[scale=0.3,angle=90]{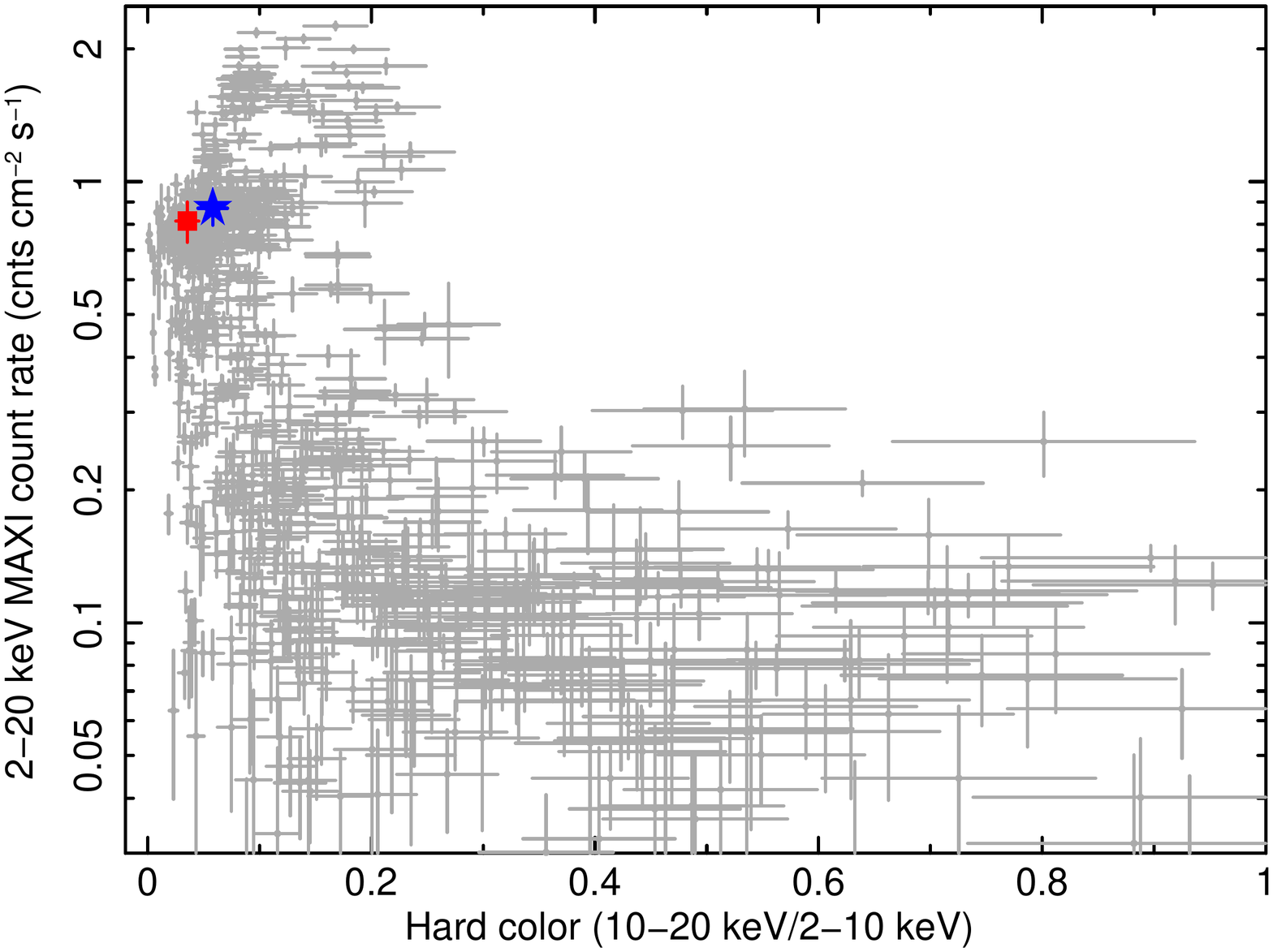}
\includegraphics[scale=0.3]{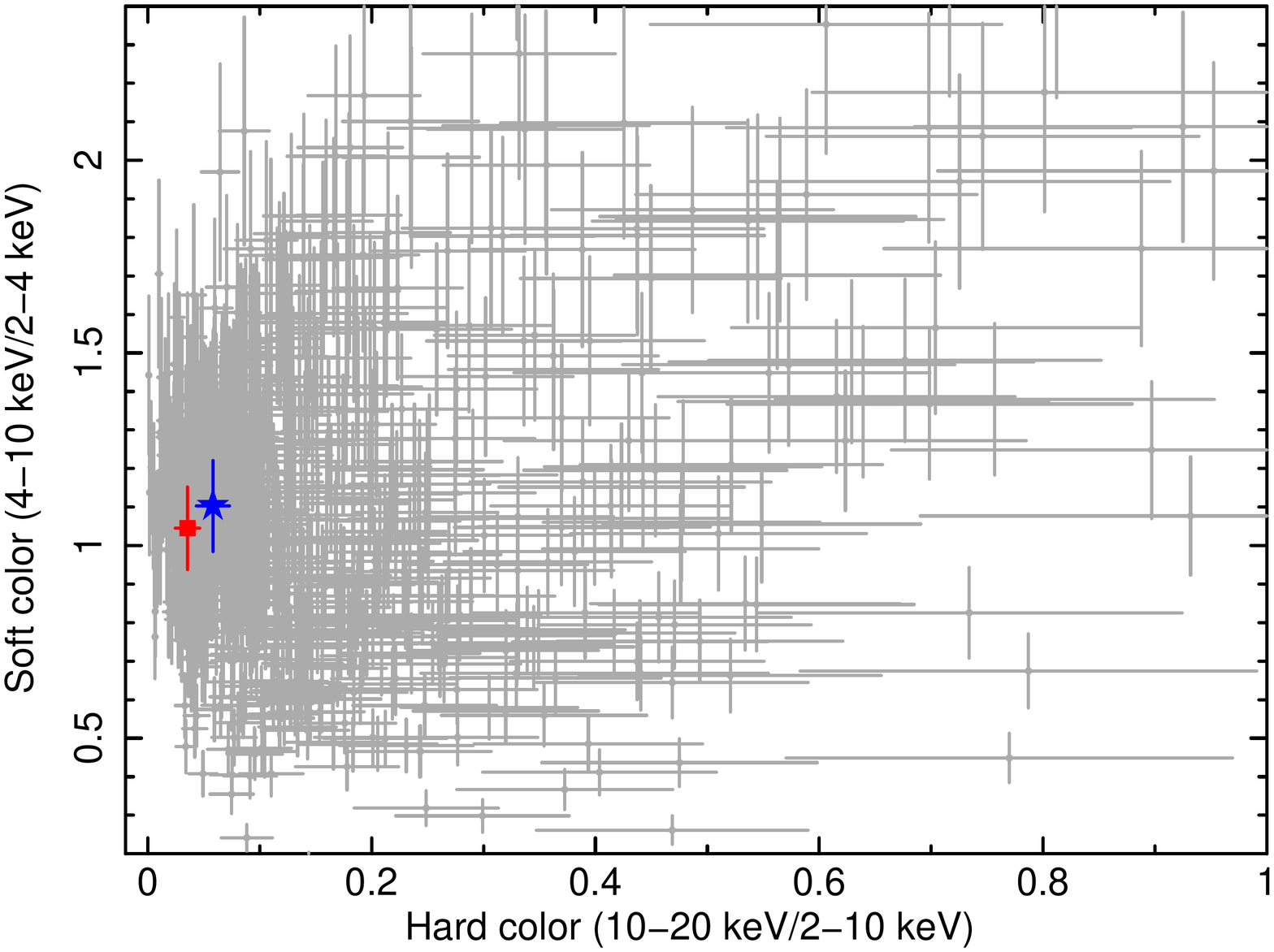}
\caption{Left panel shows the hardness-intensity diagram (HID), while the right panel shows the color-color diagram (CCD). Both HID and CCD are plotted using one-day averaged \maxi{} lightcurve of last three major outbursts from 4U 1630--47. No significant difference in the hard color or soft color value is observed during the \chan{} and \asat{} observations marked by a blue star and a red square respectively (see Section~\ref{state}).
\label{hidcc}}
\end{center}
\end{figure}

\section{Observations and data reduction}\label{Observations}

\subsection{AstroSat data reduction}\label{AstroSat}

\asat{} continuously observed 4U 1630--47 between 01 October 2016 09:16:32 and 02 October 2016 16:43:39 covering 15 consecutive satellite orbits. The total observation duration is 94.6 ks. For the broadband spectroscopic purpose, we use simultaneous observations from the Soft X-ray focusing Telescope (SXT) and Large Area X-ray Proportional Counter (LAXPC) instruments.

SXT is a focusing telescope with cooled CCD camera that can perform X-ray imaging and medium resolution spectroscopy in the 0.3-8.0 keV energy range \citep{si16,si17}. Level-1 Photon Counting (PC) mode data along with the SXT calibration database is processed through a pipeline software ({\tt AS1SXTLevel2-1.4a}; release date: 06 December 2017) to produce level-2 event files, and then a good time interval (GTI) correcter and SXT event merger script\footnote{\label{note1}\url {http://www.tifr.res.in/~astrosat\_sxt/page1\_data\_analysis.php}} are used to create a merged event file using all the clean events from different orbits with the corrected exposure time. We use {\textsc XSELECT V2.4e} in {\textsc HEASOFT 6.24} to extract lightcurves and spectra using source regions between 1 and 13 arcmin. An off-axis auxiliary response file (ARF) appropriate for the source location on the CCD is generated from the provided on-axis ARF using {\tt sxtmkarf} tool\footnoteref{note1}. A blank sky SXT spectrum, provided by the SXT team\footnoteref{note1} is used as the background spectrum. As suggested, while fitting the SXT spectrum we use the {\tt gain} command that modifies the response file gain linearly. The slope of the linear fit is fixed to 1, and the offset is free to vary. 

LAXPC consists of three large area ($\sim$6000 cm$^2$), almost identical but independent, X-ray proportional counters ({\tt LAXPC10, LAXPC20 and LAXPC30}) having absolute time resolution of 10 $\mu$s in the energy range 3.0-80.0 keV \citep{ya16a, ya16b, an17, ag17}. Owing to the high time resolution and high efficiency in hard X-rays, the LAXPC demonstrates remarkable capabilities in spectro-timing analysis of X-ray binaries like GRS 1915+105 \citep{ya16a}, Cyg X-1 \citep{mi16}, 4U 1728-34 \citep{ve17} and Cyg X-3 \citep{pa18,pa17}. Event mode data from LAXPC were acquired in 1024 channels and analyzed using the LAXPC software\footnote{\label{note2}\url {http://www.tifr.res.in/~astrosat\_laxpc/LaxpcSoft.html}}. Details of the response matrix computation and the generation of background spectra based on sky background model can be found in \citet{an17}. Due to the gas leakage issue in {\tt LAXPC30}, we do not include its spectra for further analysis. Since the energy spectra during the soft state are dominated by the LAXPC background above 23 keV, we consider the energy range of 4-23 keV for spectral analysis. 

\begin{figure*}
\begin{center}
\includegraphics[scale=0.3,angle=90]{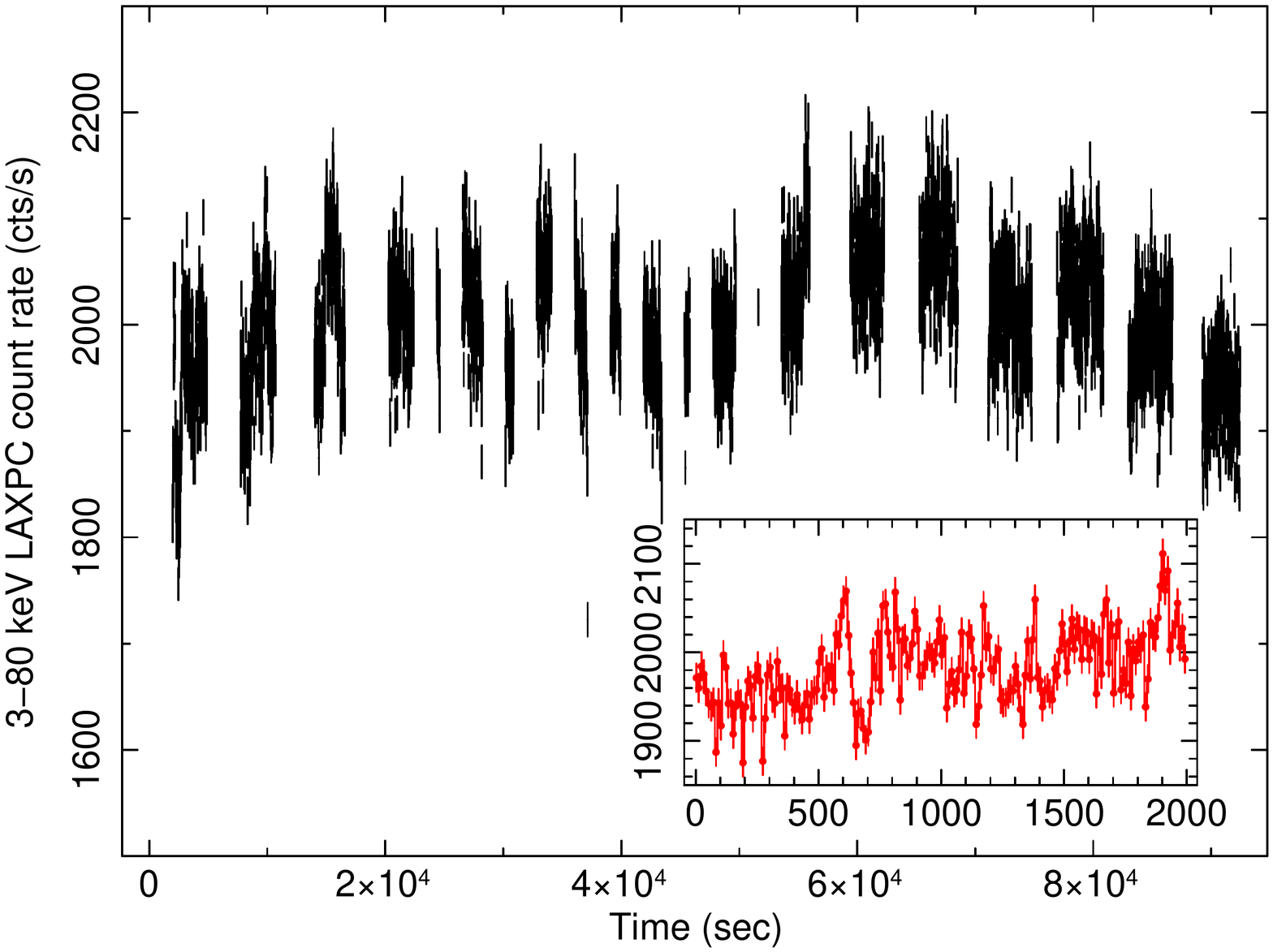}
\includegraphics[scale=0.3,angle=90]{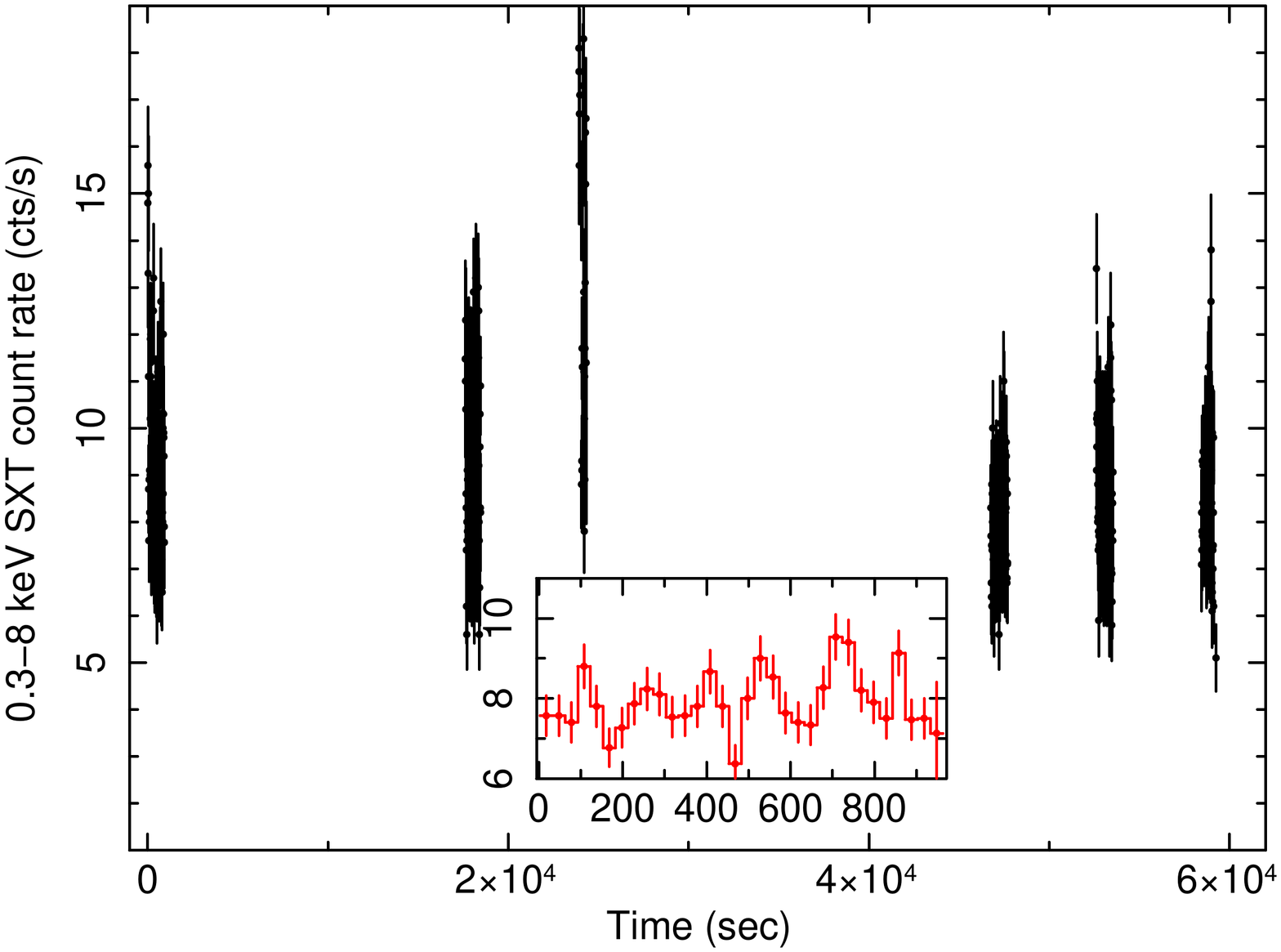}
\caption{Left panel shows the 3-80 keV background-subtracted lightcurve of 4U 1630--47 combining all three \asat{}/LAXPC units. Right panel shows 0.3-8.0 keV \asat{}/SXT lightcurve. Time bin size for both lightcurves is 10 s (see Section~\ref{state}). 2000 sec and 1000 sec sections of LAXPC and SXT lightcurves respectively are shown in the inset.}  
\label{asatlight}
\end{center}
\end{figure*}

\subsection{Chandra data reduction}\label{Chandra}

During the peak of the outburst, \chan{} observed 4U 1630--47 starting from 21 October 2016 01:24:58 (Observation ID: 19904; PI: Sudip Bhattacharyya) with an effective exposure of 30.93 ks. The observation was performed using the high-energy transmission grating spectrometer (HETGS) instrument which disperses the incident X-ray photons onto the Advanced CCD Imaging Spectrometer spectroscopic array (ACIS-S). To avoid the photon pile-up issue in the CCD, caused by the high soft unabsorbed X-ray flux of $\sim$300 mCrab in 0.2-10 keV during the present observation of 4U 1630--47, the ACIS-S array was operated in continuous clocking (CC-GRADED) data mode which activates fast frame transfer and reduces the frame accumulation time from 3.2 s to 2.85 ms. In our analysis, we determine the location of the zeroth order and owing to the highest number of events, we extract and use the first order grating spectra only.

Data reduction was accomplished using the Chandra Interactive Analysis of Observation software (CIAO version 4.10; \citet{fr06}). The Calibration Database version 4.7.8 is used. Time-averaged first-order High Energy Grating and Medium Energy Grating (MEG) spectra are extracted from the Level-2 event file. Redistribution matrix files (rmfs) and ancillary response files (arfs) are generated using the tool {\tt mkgrmf} and {\tt mkgarf} respectively. To obtain a reasonable number of energy spectral bins for the continuum spectral modeling, we group every eight spectral channels into one. All spectral analyses were conducted using {\tt XSpec version 12.10.0c}. All errors quoted in this paper are 1$\sigma$ errors, unless mentioned otherwise. 

The HEG effective area below 1 keV is very low due to soft X-ray absorption in its polyimide structure and due to truncation of the dispersion by the CCD array \citep{ca05}. During the \chan{} CC mode observations of bright and absorbed sources, a diffraction scattering halo is observed which significantly dominates the continuum spectra below 2 keV\footnote{\url {http://cxc.harvard.edu/cal/Acis/Cal\_prods/ccmode/ccmode\_final\_doc02.pdf}} and acts as a background for the source spectrum.  Therefore, considering best HEG spectral calibrations during the CC mode observation, we use the energy range of 2.0-8.0 keV for spectral analysis.

 \section{Timing analysis and spectral state determination}\label{state}
 
 \subsection{\maxi{} ans \swift{}/BAT view of the outburst}

To understand the nature of the 2016 X-ray outburst from 4U 1630--47, we analyze \maxi{} one-day averaged lightcurves in different energy ranges, as shown in Figure \ref{maxilight}. The top three panels show the 2--4 keV, 4--10 keV and 10--20 keV lightcurves covering the entire outburst which lasted for $\sim$150 days, and is significantly visible only in 2-4 keV ($\sim$200 mCrab at the peak) and 4-10 keV ($\sim$ 300 mCrab at the peak) energy ranges but not visible in the 10-20 keV energy range during the outburst duration. Therefore, the hard X-ray flux is not high enough to be detected with \maxi{}. To confirm and check the hard X-ray behavior further, we plot the 15-50 keV 1-day averaged \swift{}/BAT \citep{kr13} lightcurve of 4U 1630--47 during the same outburst. The source is not detected with the \swift{}/BAT during first 90 days of the outburst although an increase in the BAT count rate is observed during the decay phase of the outburst. Therefore, the non-detection of 4U 1630--47 in hard X-rays at the beginning of the outburst implies that the source went into a soft X-ray outburst. The times of observations from \chan{} and \asat{} are shown by vertical lines in Figure \ref{maxilight}. We may note that, although there is a gap of $\sim$20 days, both observations were taken during the spectral state when hard X-rays are not detected with \swift{}/BAT.  

\subsection{HID and CCD analysis}

To show both of our observations belong to the canonical soft state and not the intermediate state, we plot hardness-intensity diagram (HID) and color color diagram (CCD) which are shown in the left and right panels of Figure \ref{hidcc} respectively. The hard color is defined as the ratio of background-subtracted count rate between 10--20 keV and 2--10 keV, while the soft color is defined as the ratio of background-subtracted count rate between 4--10 keV and 2--4 keV. To plot HID and CCD, we use one-day averaged \maxi{} data \citep{ma09} of the last three major outbursts from 4U 1630--47 in 2016, 2015 and 2012. \maxi{} hard and soft colors during the current \chan{} and \asat{} observations are shown by a blue star and a red square respectively. From both HID and CCD, we may note that the hard color and soft color are not significantly different during \chan{} and \asat{} observations. The similarities in the CCD and HID implies, although these observations are separated by $\sim$20 days, the spectral states are same. If we compare the HID of 4U 1630--47 from \maxi{} with that from \xte{} as presented by \citet{re06}, we find that the source spent most of the time in the canonical high soft state during outbursts (close to the lowest hard color or the leftmost section of the `q' diagram), and not in the intermediate state (horizontal tracks of the `q' diagram). Therefore, both our \chan{} and \asat{} observations were performed during the canonical high soft state.

\begin{figure}
\begin{center}
\includegraphics[scale=0.4]{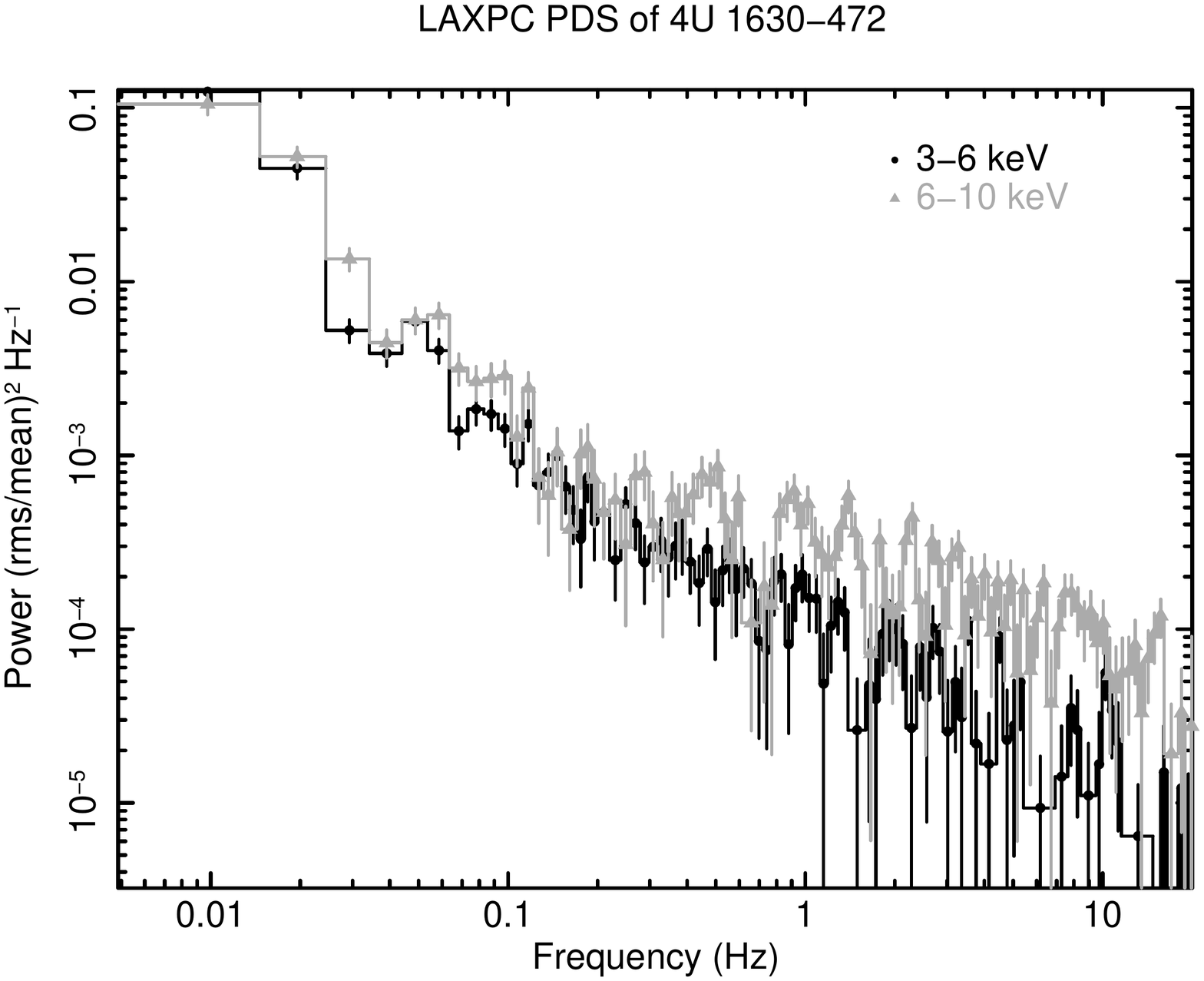}
\caption{Deadtime-corrected, Poisson noise subtracted and background corrected power density spectra (PDS) from \asat{}/LAXPC in the energy ranges 3--6 keV and 6--10 keV are shown in the frequency range of 5 mHz to 20 Hz. PDS are featureless and weak, bearing the signature of the high soft state (see Section~\ref{state}).}
\label{pds}
\end{center}
\end{figure} 

\subsection{\asat{} observations}

The \asat{}/LAXPC background-subtracted lightcurves combining three LAXPC units in the 3--80 keV light curve is shown in the left panel of Figure \ref{asatlight}, while the 0.3--8 keV \asat{}/SXT light curve is shown in the right panel. Time bin size of both lightcurves is 10 s. No strong variation in count rate is observed. To check for the presence of any quasi-periodic oscillations (QPOs) in the light curve, we plot power density spectra (PDS) in Figure \ref{pds}. PDS are extracted using LAXPC data in the energy ranges 3--6 keV and 6--10 keV. PDS are deadtime-corrected, Poisson-noise subtracted and are also corrected for background. PDS are dominated by a red noise component, and no significant power is observed above $\sim 1$ Hz at the significance level of 2$\sigma$. The total rms power integrated over $0.1-10$ Hz is less than 3\% in 3--80 keV. These PDS properties are similar to the characteristics of the canonical high soft state \citep[e.g., ][]{re06}.

\subsection{Radio observations}

4U 1630--47 was also observed in radio wavelength using Australian Telescope Compact Array (ATCA) on 28 September and 21 October 2016 which was simultaneous with the \chan{} observation. Both observations were taken at 5.5 GHz and 9 GHz. At these frequencies, no source at the position of 4U 1630--47 was detected at the 3$\sigma$ rms noise level of 17 $\mu$Jy and 16 $\mu$Jy respectively on 21 October 2016. At the similar rms level, no source was detected on 28 September 2016. The radio non-detection during our observation is in agreement with the high soft spectral state of the source \citep{re06}.
     
\section{Spectral analysis and results}\label{Spectral} 

As described in Section~\ref{state}, using \maxi{}, \swift{}/BAT and \asat{} data, and from light curves, HID, CCD and PDS, we conclude that both \chan{} and \asat{} observations were performed in the high soft state. Therefore, it is meaningful to do combined \chan{} and \asat{} spectral analysis. But, to be cautious, first we carry out \chan{} grating spectral analysis and \asat{}/SXT+LAXPC joint spectral analysis separately. 

\subsection{\chan{} grating spectral analysis}\label{grating}

\begin{figure*}
\begin{center}
\includegraphics[scale=0.3]{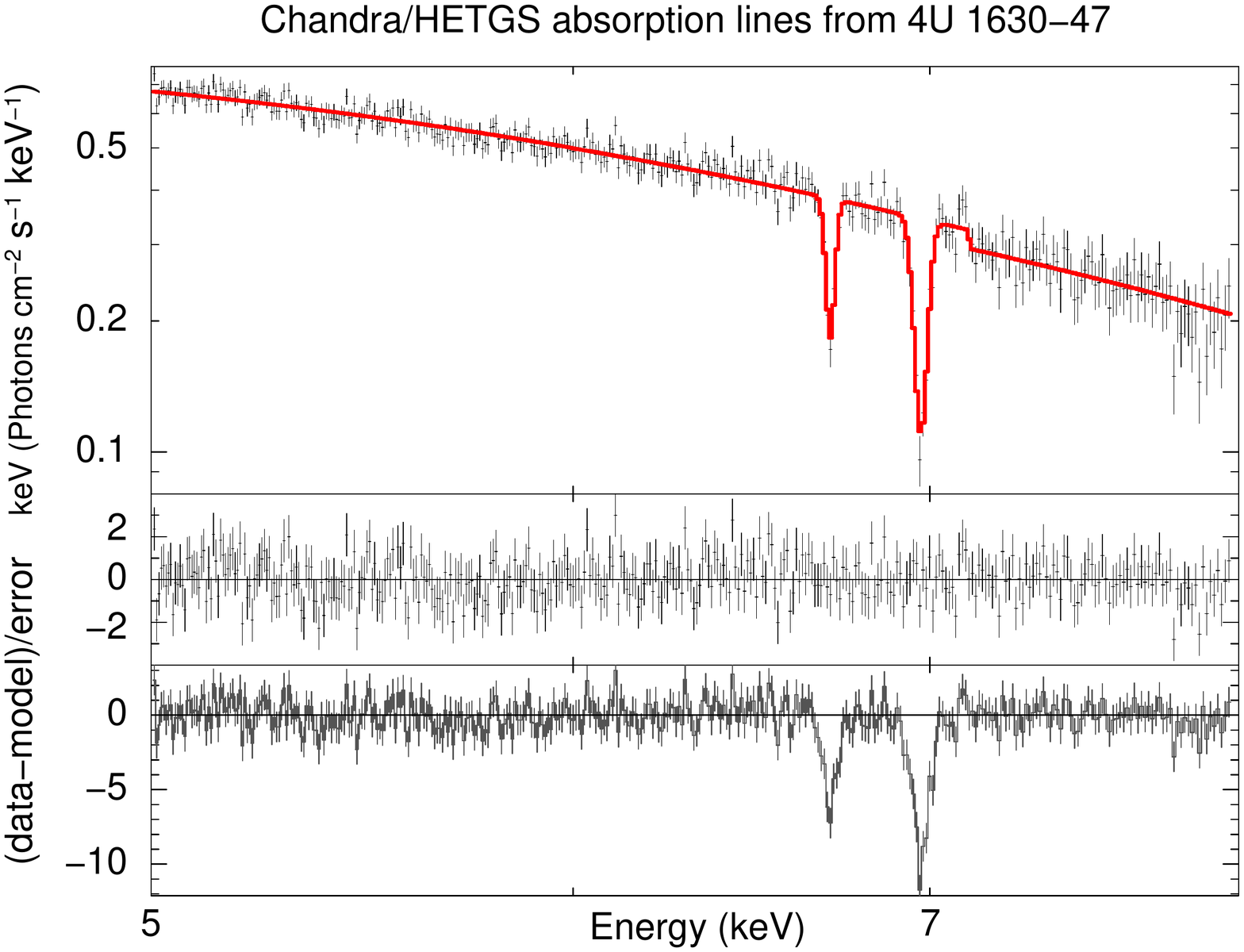}
\includegraphics[scale=0.3]{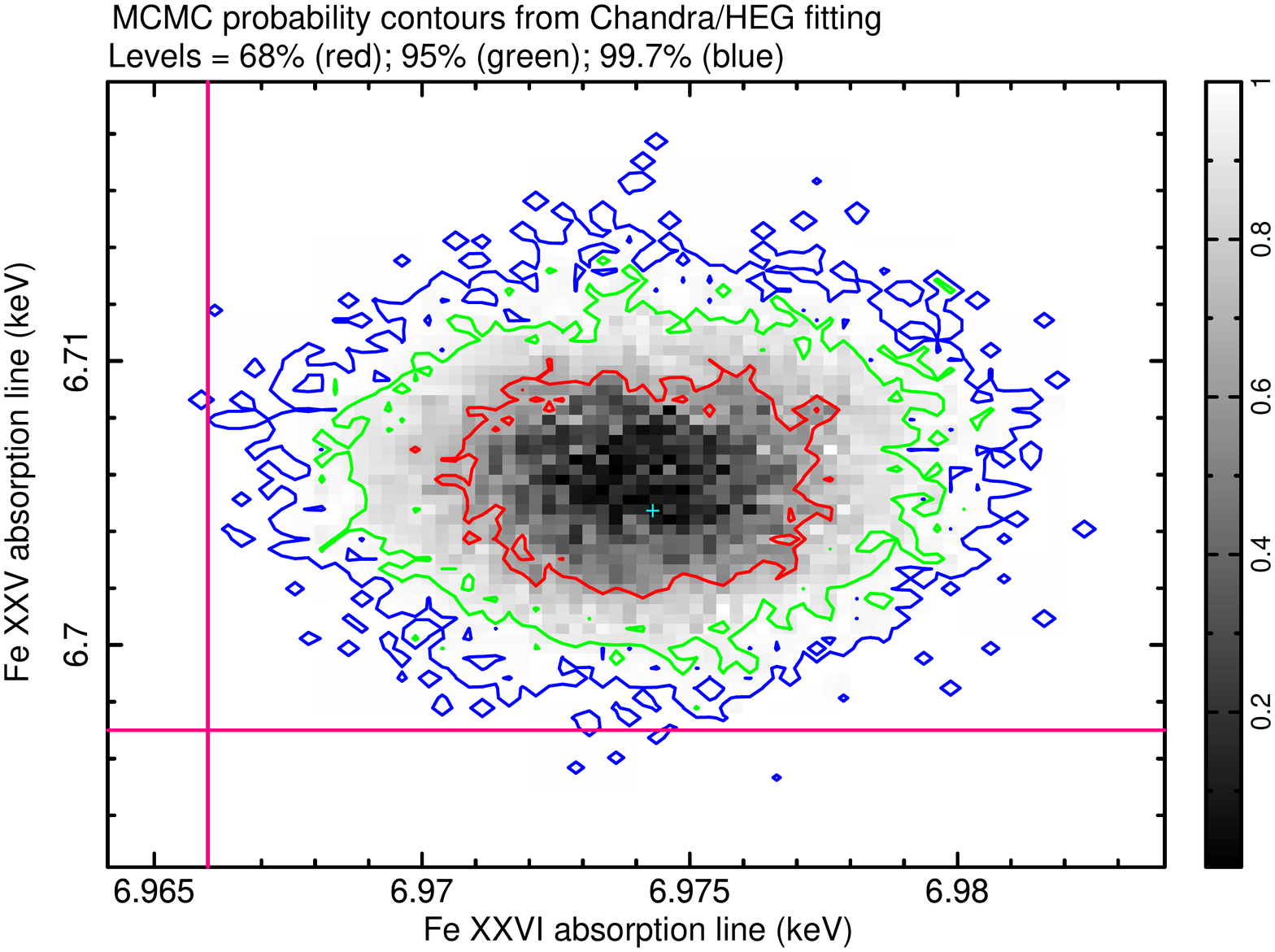}
\caption{Absorption lines from \chan{} grating spectra. Left panel shows zoomed, unbinned first order \chan{}/HEG spectrum of 4U 1630--47, when fitted with an absorbed single disk-blackbody model, along with the residual. Two strong absorption lines visible at $\sim 6.7$ keV and $\sim 6.97$ keV which are due to the ionized Fe XXV and Fe XXVI absorption features are fitted with Gaussian absorption profiles. Bottom left panel shows residual without the line absorption models. Right panel shows 1$\sigma$, 2$\sigma$ and 3$\sigma$ integrated contours from marginal probability distribution of Fe XXV absorption line energy versus Fe XXVI absorption line energies as obtained from \chan{}/HEG spectral analysis. The horizontal and the vertical magenta lines denote the rest frame energies of Fe XXV and Fe XXVI lines at 6.697 keV and 6.966 keV respectively. A blueshift of $\sim$0.0012c is detected for both absorption lines with at least 3$\sigma$ significance.}
\label{abs}
\end{center}
\end{figure*}

To understand characteristics of the \chan{}/HETGS spectra, we fit the binned, first-order high energy gratings (HEG) spectrum with a thermal disk blackbody model along with the line-of-sight absorption model {\tt phabs} in {\tt XSpec}. The abundance is set to {\tt aspl} from \citet{as09}. Due to calibration issues, we do not use spectra from the medium energy gratings. Motivated by the previous work by \citet{ki14}, which demonstrated the presence of a fast-spinning black hole, we use a multi-temperature, relativistic, disk blackbody model {\tt kerrbb} for a thin, steady state, general relativistic accretion disk around a Kerr black hole \citep{li05}. This model fits the broadband continuum well. The residual of the fitting shows two strong absorption features. Due to symmetric shapes of the absorption profiles, we use a Gaussian absorption model to account for the two strong absorption lines at $\sim$6.7 keV and $\sim$6.97 keV.
In our {\tt XSpec} model {\tt phabs*(kerrbb+gabs+gabs)}, we fix the inclination at $i = 64^\circ$, black hole mass at $M = 10.1 M_\odot$ and the distance at $D = 10$~kpc, based on our discussion in Section~\ref{Introduction}. The torque at the inner disk boundary is assumed to be zero. The effect of limb-darkening and self-irradiation are also included. The HEG spectrum can be well fitted by the best-fit model ($\chi^2$/dof $= 250/252 (0.99)$). Best-fit spectral parameters are given in Table \ref{fitparm}, and the fitted spectra along with residuals are shown in the top left panel of Figure \ref{spectra}. 
However, to determine the nature and characteristics of absorption lines, we use the original unbinned spectrum to retain the original spectral resolution. The left panel of Figure \ref{abs} shows a zoomed portion of the unbinned, continuum-fitted spectra between 5 and 8 keV, where two absorption features at $\sim$6.7 keV and $\sim$6.97 keV are observed from the residual in the bottom panel. These line energies correspond to Fe XXV and Fe XXVI absorption lines at the rest energy of 6.697 keV and 6.966 keV respectively \citep{bi05}.  
Best-fit Fe XXV and Fe XXVI absorption line energies are found to be 6.705$^{+0.002}_{-0.002}$ keV and 6.974$^{+0.004}_{-0.003}$ keV, which are blueshifted from their rest frame energies by an amount which corresponds to an outflow velocity of $\sim$0.0012c. The blueshift is measured with 3$\sigma$ significance, and 1$\sigma$, 2$\sigma$ and 3$\sigma$ contours are shown in the right panel of Figure \ref{abs} along with the rest frame energies marked by a vertical and a horizontal line. From the shift in line energies, we estimate the wind velocity is 366 $\pm$ 56 km/s.   
Significances are calculated based on Markov Chain Monte Carlo (MCMC) simulations of fitted parameters, which are described later. Best-fit black hole spin parameter and spectral hardening factor are 0.928$^{+0.029}_{-0.035}$ and 1.56$^{+0.08}_{-0.09}$ respectively. The spectral hardening factor, defined as a multiplicative factor to relate the effective temperature and the color/observed temperature of the accretion disk \citep{sh95}, is low possibly due to the soft, thermal nature of the spectra \citep{sa13}.

\subsection{\asat{} spectral analysis}\label{astrosat}

To check the \chan{} spectral fitting consistency and to examine the hard X-ray behavior above 8 keV, we jointly fit the simultaneous \asat{}/SXT and \asat{}/LAXPC spectra in the energy range of $0.5-23.0$ keV. By fitting with the above mentioned continuum model, we find a significant residual above 16 keV and the fit is unacceptable ($\chi^2$/dof = 624/214). To account this, we use a convolving Comptonization model {\tt simpl} in {\tt XSpec}. {\tt simpl} is an empirical but self-consistent model of Comptonization, in which a fraction of the thermal photons from a thin disk works as an input seed spectrum \citep{St09}. With this modification, the spectrum is fitted well with the $\chi^2$/dof = 209/212 (0.99). \asat{}/SXT and \asat{}/LAXPC spectra are not sensitive enough to model two absorption features similar to that shown by the \chan{}/HEG spectrum. Therefore, we fix absorption line parameters to that from \chan{}/HEG fitting.  
Best fit spectra along with the residual is shown in the top right panel of Figure \ref{spectra}, while the best-fit parameter values are given in Table \ref{fitparm}. 
Since spectral hardening factor is not constrained, we fix it to 1.57 (see Section~\ref{grating}). Best-fit black hole spin parameter is found to be 0.913$^{+0.012}_{-0.028}$ (3$\sigma$ errors), which is close to that estimated from \chan{}/HEG spectral fitting. From Table \ref{fitparm}, we may note that {\tt kerrbb} best-fit parameters for \asat{} and \chan{} data are close to each other. 
A direct implication of such similarity in spectral parameters is that spectral state do not evolve between the \asat{} and \chan{} observation periods. This motivates us to perform \chan{} and \asat{} joint spectral analysis.

\subsection{\chan{} + \asat{} joint spectral analysis}\label{chanasat}

\begin{figure*}
\begin{center}
\includegraphics[scale=0.3]{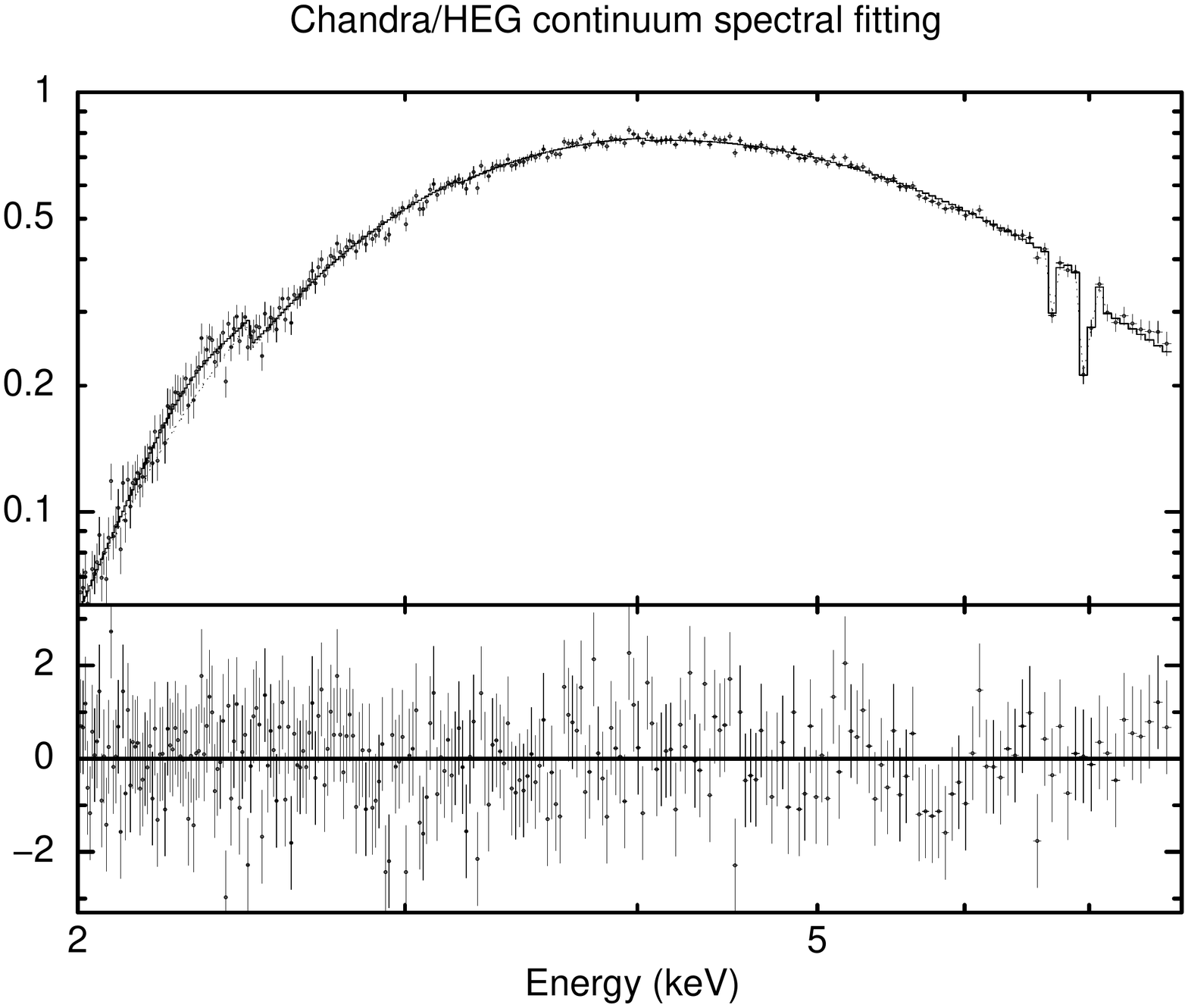}
\includegraphics[scale=0.3]{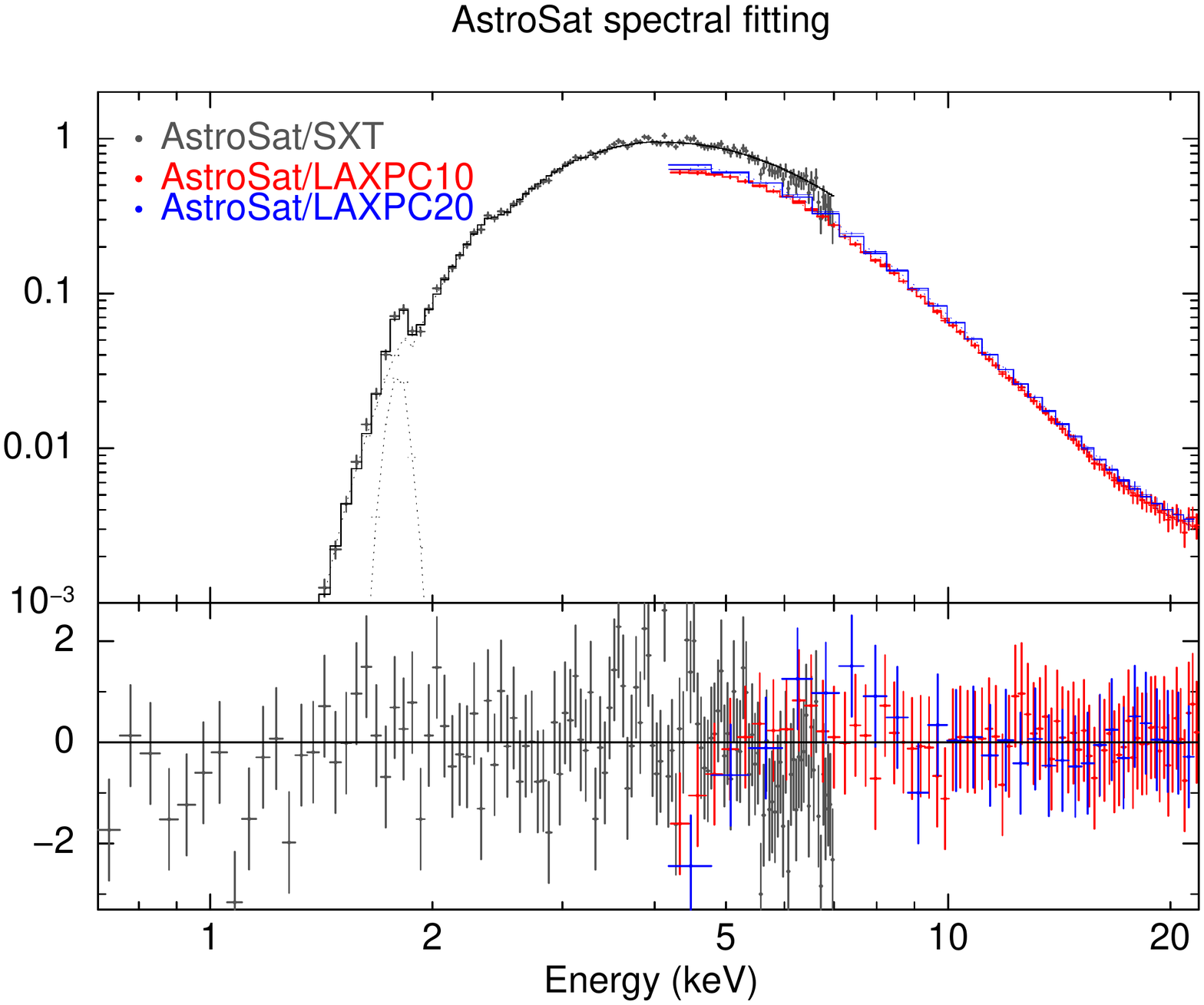}
\includegraphics[scale=0.3]{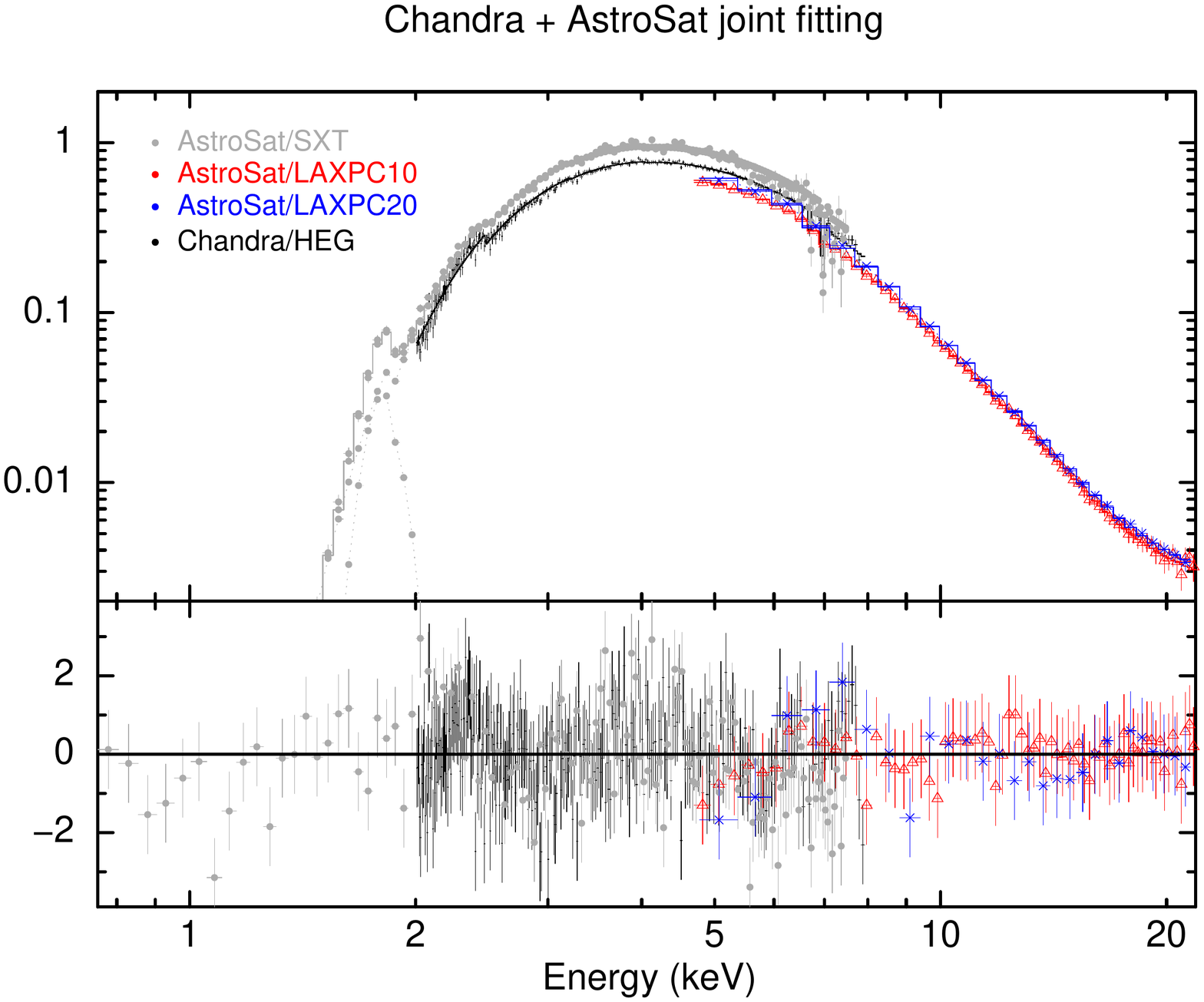}
\includegraphics[scale=0.3]{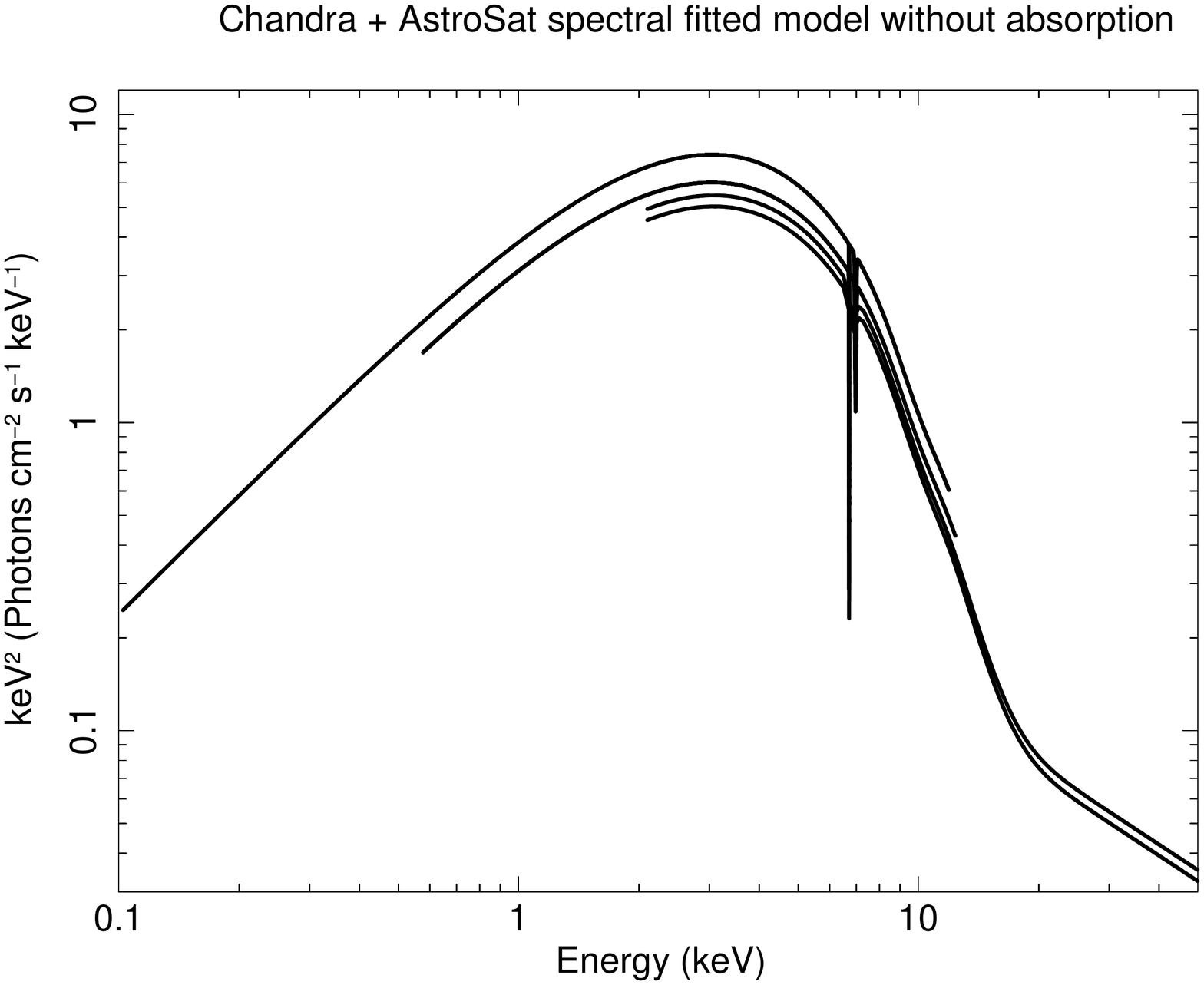}
\caption{Top left panel shows the \chan{}/HEG spectral fitting (with a spectral binsize of 8 channels), and the residual using an absorbed, relativistic, disk-blackbody model {\tt kerrbb} along with Gaussian absorption features. With this model an additional convolution Comptonization model {\tt simpl} is used to fit \asat{}/SXT, \asat{}/LAXPC10, \asat{}/LAXPC20 joint spectra, which is shown in the top right panel along with the residual. Bottom left panel shows the joint fitting of \chan{}/HEG, \asat{}/SXT+LAXPC10+LAXPC20 spectra with the same continuum model as the previous one. Bottom right panel shows the unfolded model spectra without the absorption in the unit of E*E*f(E). The model spectrum peaks at $\sim$3-4 keV. }  
\label{spectra}
\end{center}
\end{figure*}

We perform \chan{}/HEG and \asat{}/SXT+LAXPC joint spectral analysis to check if spectral parameters from \chan{} and \asat{} individual spectral modelling match with \chan{}+\asat{} joint fitting. We use the continuum model similar to that used for SXT+LAXPC joint spectral analysis discussed above (see Section~\ref{astrosat}). Absorption line parameters for the SXT spectrum are tied to that with the HEG. Best fit spectra ($\chi^2$/dof = 507/477) along with the residual are shown in the bottom left panel of Figure \ref{spectra}, while best-fit parameter values are given in Table \ref{fitparm}. The unfolded best-fit model spectra excluding the absorption is shown in the bottom right panel of Figure \ref{spectra}.  
Best-fit black hole spin parameter and spectral hardening factor with 3$\sigma$ errors are 0.924$^{+0.016}_{-0.007}$ and 1.56$^{+0.06}_{-0.09}$ respectively. All best-fit parameters are found similar to those estimated from separate \asat{} and \chan{} spectral analyses. 

\begin{table*}
 \centering
 \caption{Best-fit parameters using individual as well as joint \chan{} and \asat{} spectral fitting during the high soft state of 4U 1630--47}
\begin{center}
\scalebox{0.92}{%
\begin{tabular}{ccccccccc}
\hline 
model & parameters & \chan{}  & \asat{} & \chan{}+\asat{} \\
 & & HEG  & SXT+LAXPC & SXT+HEG+LAXPC \\
\hline
constant & & -- & SXT 1(fixed) & Chandra 1(fixed) \\
 & & & LAXPC10 0.95$^{+0.04}_{-0.03}$ & LAXPC10 0.92$^{+0.12}_{-0.11}$ \\
 & & & LAXPC20 1.01$^{+0.04}_{-0.08}$ & LAXPC20 0.96$^{+0.05}_{-0.03}$ \\
 & & &  & SXT 1.13$^{+0.09}_{-0.11}$ \\   
\hline
{\tt phabs} & N$_{H}$ (10$^{22}$ cm$^{-2}$) & 11.8$^{+1.86}_{-1.28}$  & 11.2$^{+1.3}_{-2.4}$ & 11.5$^{+0.7}_{-1.1}$ \\
{\tt kerrbb} & $\dot{M}$ (10$^{18}$ gm/s) & 1.49$^{+0.07}_{-0.06}$ & 2.17$^{+0.11}_{-0.10}$ & 1.51$^{+0.19}_{-0.11}$ \\
                      & $a_*$ (3$\sigma$) & 0.928$^{+0.029}_{-0.035}$ & 0.913$^{+0.012}_{-0.028}$ & 0.924$^{+0.016}_{-0.007}$ \\
                      & ${\it h_{d}}$ & 1.57$^{+0.08}_{-0.07}$ & 1.57(f) & 1.56$^{+0.06}_{-0.09}$ \\
{\tt simpl} & $\Gamma$ & --  & 2.79$^{+0.11}_{-0.08}$ & 2.84$^{+0.07}_{-0.08}$ \\
			 & F$_{sc}$ (\%) & --  & 3.83$^{+0.15}_{-0.22}$ & 3.33$^{+0.12}_{-0.17}$ \\
{\tt gabs}   & E$_{gabs1}$ (keV) & 6.705$^{+0.002}_{-0.002}$  & -- & 6.708$^{+0.004}_{-0.005}$ \\
                    & $\sigma_{gabs1}$ (eV) & 12.8$^{+3.4}_{-3.5}$ & -- & 11.3$^{+1.8}_{-1.3}$ \\
                   & $S_{gabs1}$ & 0.76$^{+0.11}_{-0.12}$ & -- & 0.72$^{+0.05}_{-0.07}$ \\
{\tt gabs}   & E$_{gabs2}$ (keV) & 6.974$^{+0.004}_{-0.003}$ & -- & 6.975$^{+0.007}_{-0.005}$ \\
                    & $\sigma_{gabs2}$ (eV) & 19.6$^{+2.1}_{-1.9}$  & -- & 18.1$^{+3.2}_{-2.8}$ \\
                   & $S_{gabs2}$ & 1.15$^{+0.11}_{-0.13}$  & -- & 1.49$^{+0.06}_{-0.16}$ \\                   
F$_{0.2-2}$   & (10$^{-8}$ ergs s$^{-1}$ cm$^{-2}$) & 0.81$^{+0.08}_{-0.11}$  & 1.04$^{+0.11}_{-0.14}$ & 1.05$^{+0.07}_{-0.06}$ \\
F$_{2-10}$   & (10$^{-8}$ ergs s$^{-1}$ cm$^{-2}$) & 0.92$^{+0.11}_{-0.09}$ & 1.45$^{+0.11}_{-0.09}$ & 1.33$^{+0.08}_{-0.05}$  \\
F$_{10-20}$   &  (10$^{-8}$ ergs s$^{-1}$ cm$^{-2}$) & --  & 0.15$^{+0.03}_{-0.03}$ & 0.16$^{+0.02}_{-0.02}$ \\
F$_{20-100}$   &  (10$^{-8}$ ergs s$^{-1}$ cm$^{-2}$) & --  & 0.02$^{+0.01}_{-0.01}$ & 0.03$^{+0.01}_{-0.01}$ \\
              & $\chi^2$/dof & 250/252 (0.99) & 222/219 (1.01) & 507/477 (1.06) \\
\hline
\end{tabular}}
\tablecomments{N$_{H}$ is the line-of-sight absorption column density, $\dot{M}$ is the mass accretion rate in 10$^{18}$ gm/s, $a_*$ is the black hole spin parameter, ${\it h_{d}}$ is the spectral hardening factor, $\Gamma$ is the photon powerlaw index, F$_{sc}$ is the scattered Comptonization fraction. E$_{gabs1}$, E$_{gabs2}$  and $\sigma_{gabs1}$, $\sigma_{gabs2}$ are the energies and widths of the Gaussian-shaped absorption lines in keV and eV respectively, while $S_{gabs1}$ and $S_{gabs2}$ are the strength of these absorption features. F$_{0.2-2}$, F$_{2-10}$, F$_{10-20}$ and F$_{20-80}$ are unabsorbed fluxes in the energy range $0.2-2$, $2-10$, $10-20$ and $20-100$ keV respectively in the unit of 10$^{-8}$ erg s$^{-1}$ cm$^{-2}$.  }
\end{center}
\label{fitparm}
\end{table*}

\subsection{MCMC simulations and results}\label{MCMC}

To check whether best-fit parameter values from \chan{}, \asat{} and \chan{}+\asat{} joint 
spectral analyses represent global solutions, we perform MCMC simulations of spectral parameters of all three spectral fits.
If a large number of free parameters are involved in spectral modeling (for example, our best-fit spectral model has maximum 13 free parameters including cross calibration factors), then the use of $\chi^2$ minimization technique is not always reliable for estimating the model parameters \citep{re12}. 
As an independent check, we employ the following MCMC simulations technique for validation of results obtained from our $\chi^2$ minimization method. 
As X-ray spectral counts usually follow a Poisson distribution, we replace the conventional $\chi^2$ fit statistic with the appropriate statistic for Poisson data ({\tt pgstat} in {\tt XSpec}). This assumes a Poisson distribution of source spectral counts, but a Gaussian distribution of background counts. The derivation of the profile likelihood of {\tt pgstat} is similar to that of the likelihood of C$-$statistic. With the new fit statistics, we run 5$\times$10$^5$ element chains starting from a random perturbation away from the best-fit and ignoring the first 50000 elements of the chain. The distribution of the current proposal (i.e., an assumed probability distribution for each Monte Carlo step to run the simulation) is assumed to be Gaussian with the rescaling factor of 0.001. We use the Goodman-Weare algorithm for MCMC simulations with 10 walkers. 

\begin{figure*}
\begin{center}
\includegraphics[scale=0.31]{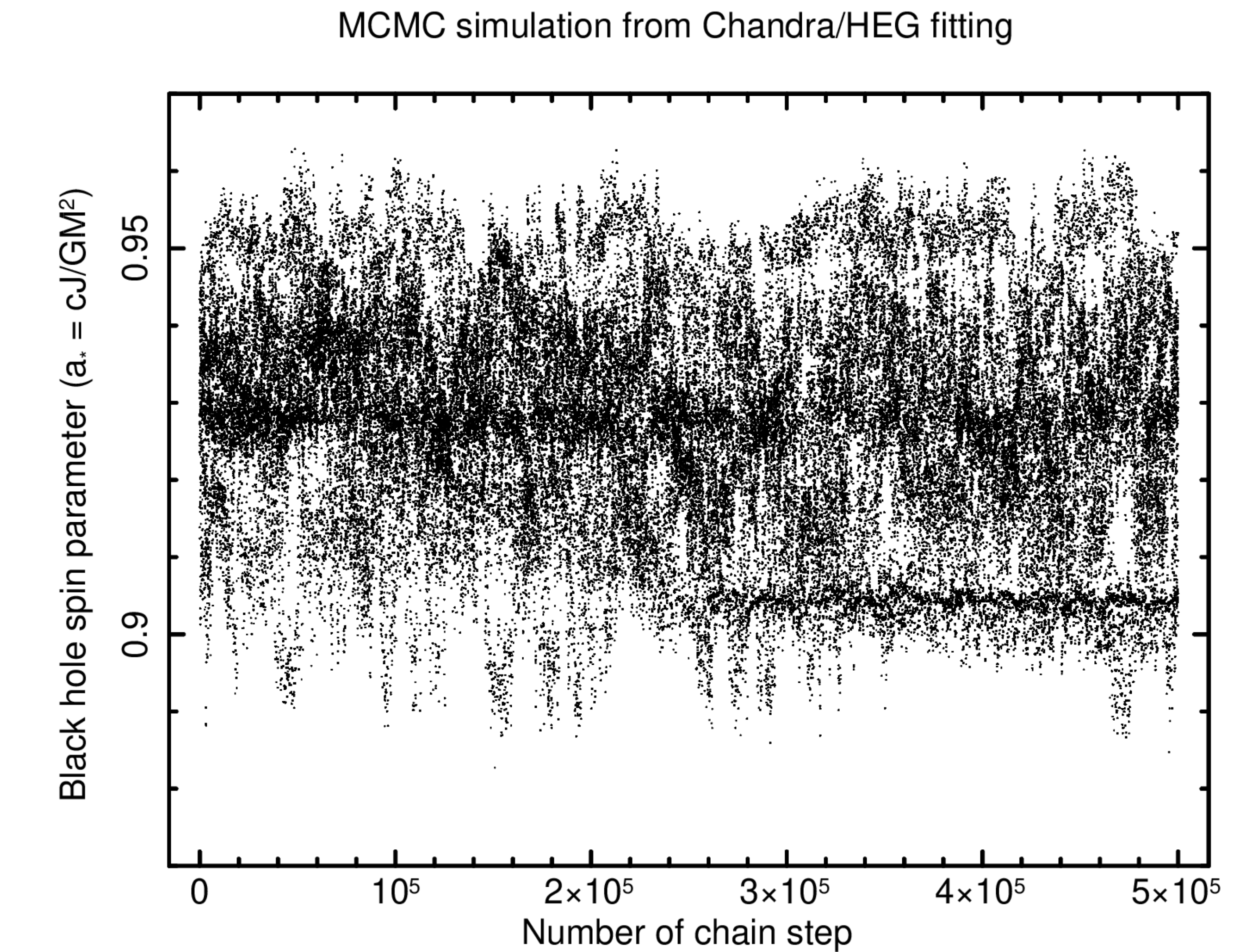}
\includegraphics[scale=0.31]{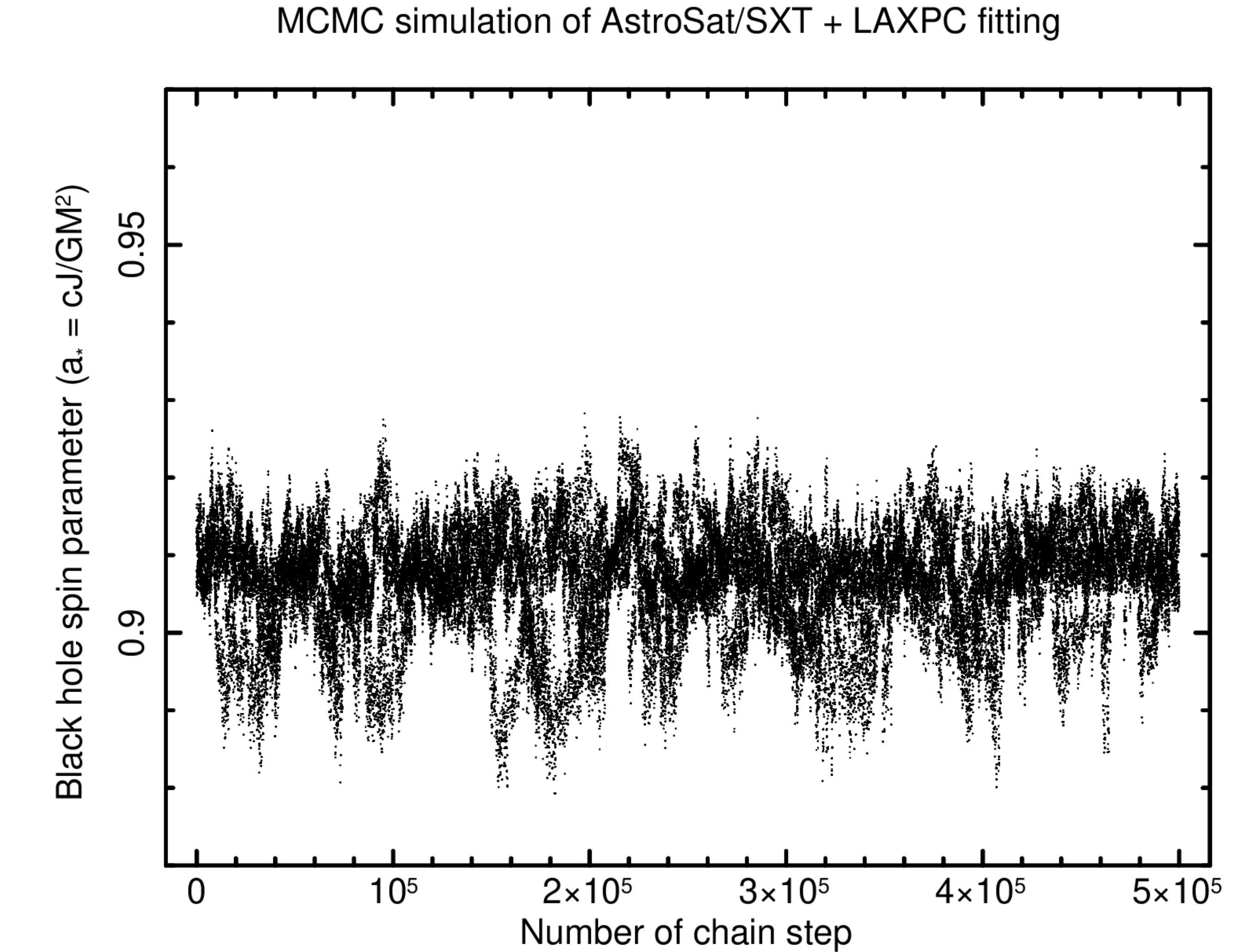}
\includegraphics[scale=0.31]{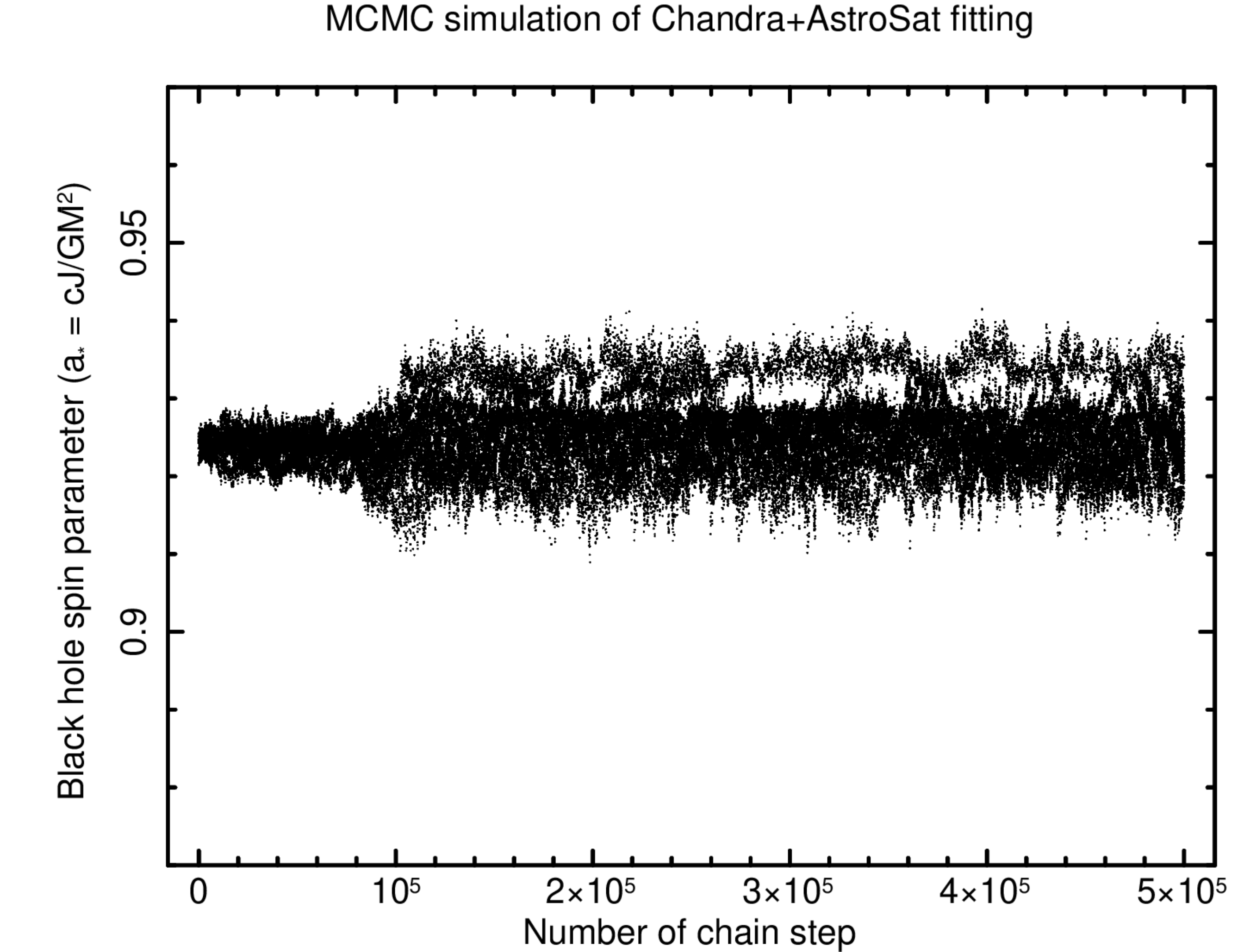}
\includegraphics[scale=0.31]{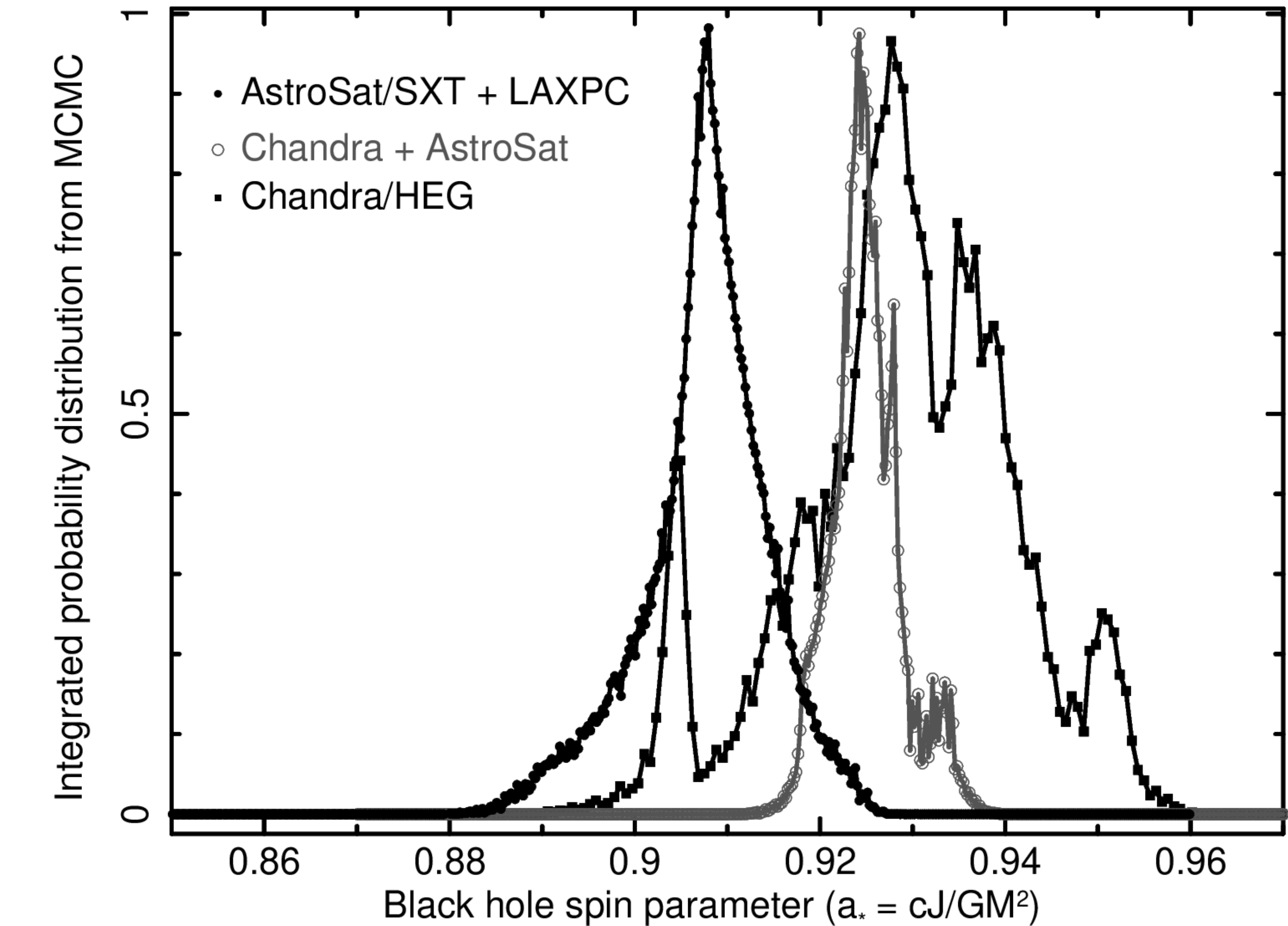}
\caption{Results from Markov Chain Monte Carlo (MCMC) simulations of fitted spectra. MCMC simulation-derived black hole spin parameter versus number of chain step used in simulations are shown for \chan{}/HEG spectral fitting (top left panel), \asat{}/SXT+\asat{}/LAXPC joint spectral fitting (top right panel) and \chan{}/HEG+\asat{}/SXT+\asat{}/LAXPC joint spectral fitting (bottom left panel). Bottom right panel shows the probability distributions of the spin parameter for three different joint spectral fits. Note that the black hole spin parameter from all three distributions is consistent with 0.92$^{+0.04}_{-0.04}$ (see Section~\ref{MCMC}).}  
\label{mcmc}
\end{center}
\end{figure*}

The top left, top right and bottom left panels of Figure \ref{mcmc} show the black hole spin parameter as a function of the number of MCMC chain steps during \chan{}/HEG, \asat{}/SXT+LAXPC and \chan{}/HEG+\asat{}/SXT+LAXPC spectral fits respectively. 
We may note that during the \chan{}, \asat{} as well as \chan{}+\asat{} joint spectral fitting the spin parameter is consistent within the range of $0.88-0.96$. The bottom right panel shows the probability distributions of the black hole spin parameter obtained from MCMC simulations for three different spectral fits. The three probability distributions have significant overlap with each other, and their 3$\sigma$ range is $0.88-0.96$. However, it is important to check how well the spectral hardening factor is constrained, since the spin parameter is known to have a dependency on the spectral hardening factor.

\begin{figure*}
\begin{center}
\includegraphics[scale=0.3]{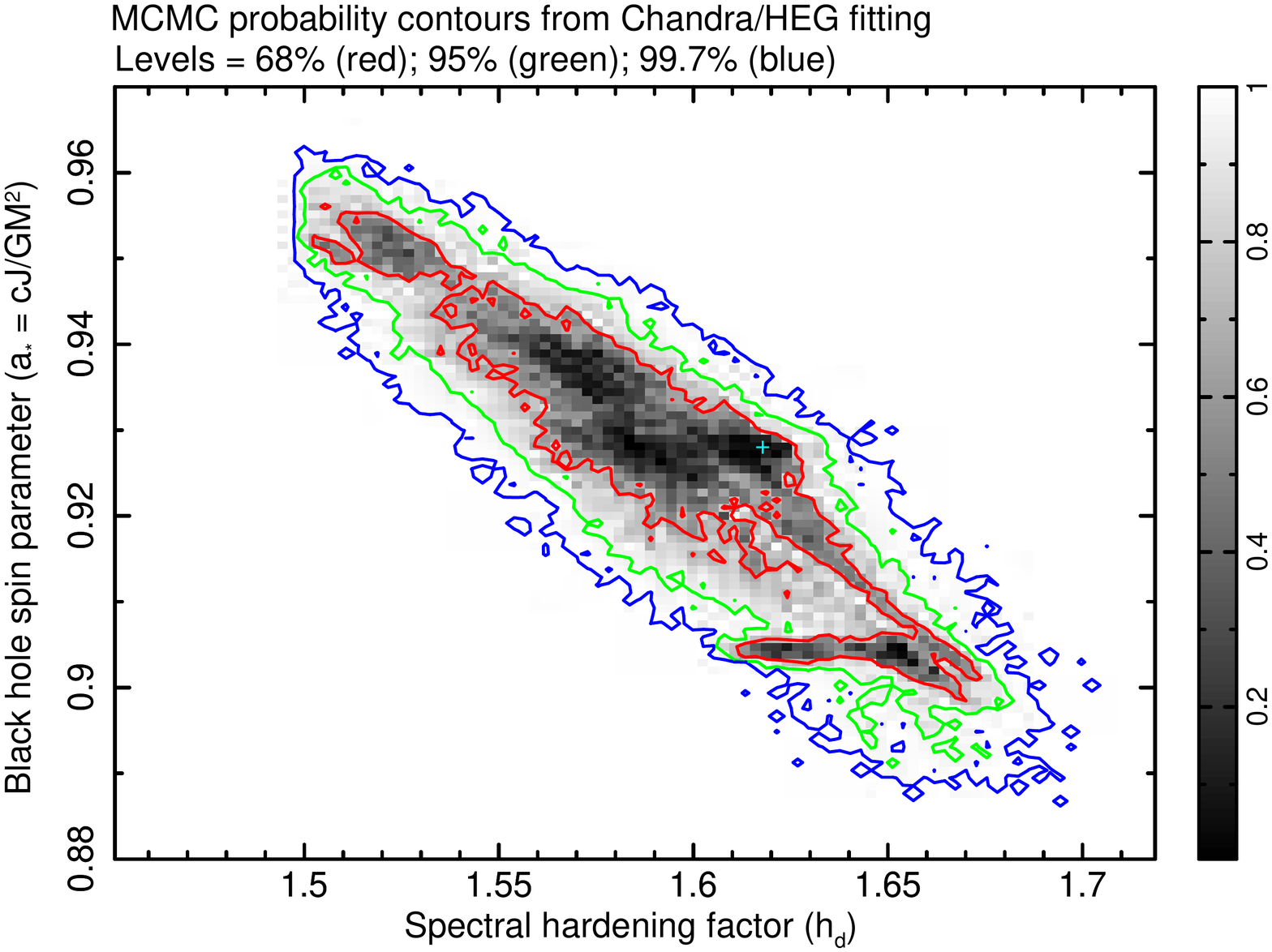}
\includegraphics[scale=0.3]{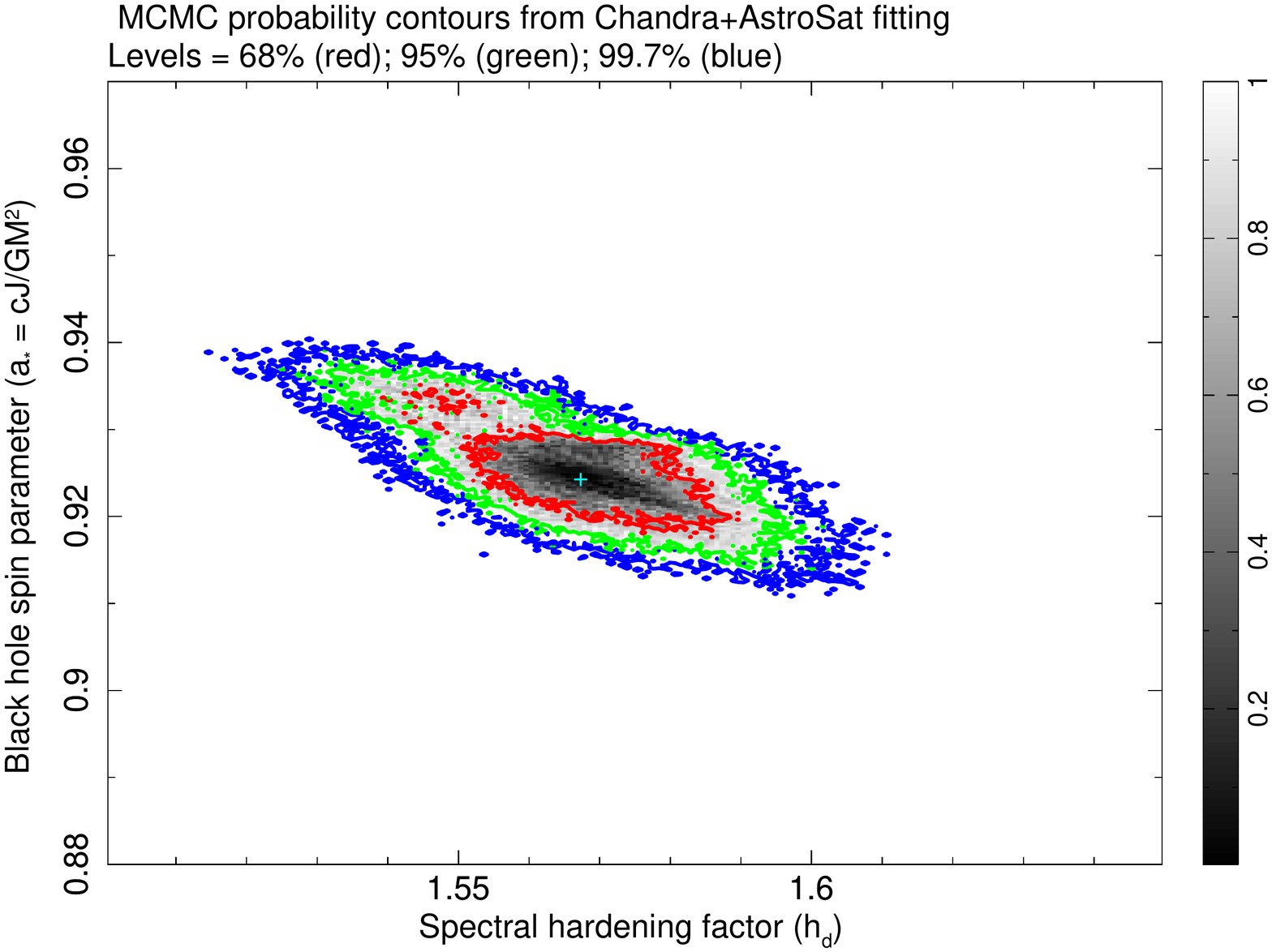}
\includegraphics[scale=0.3]{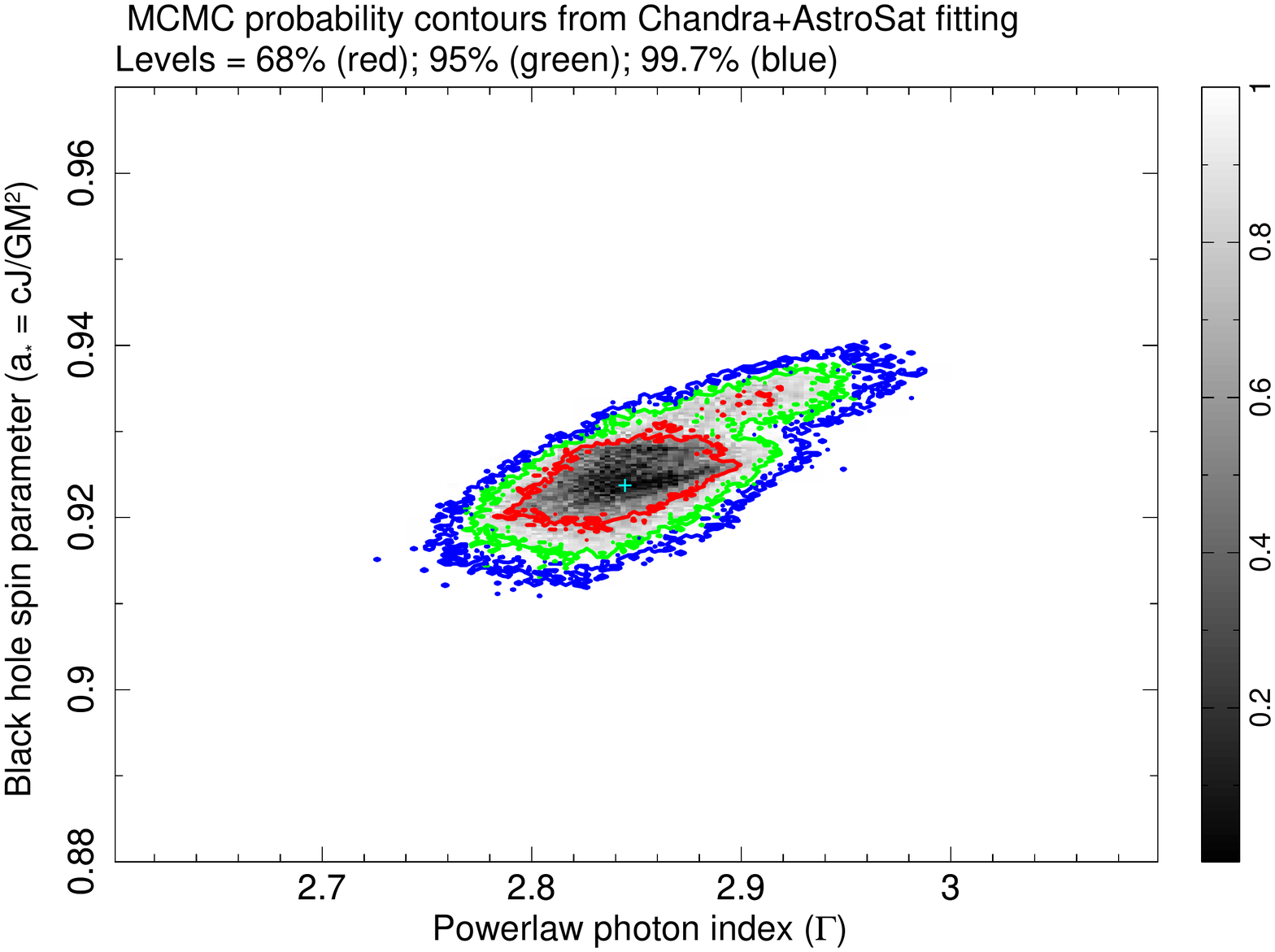}
\includegraphics[scale=0.3]{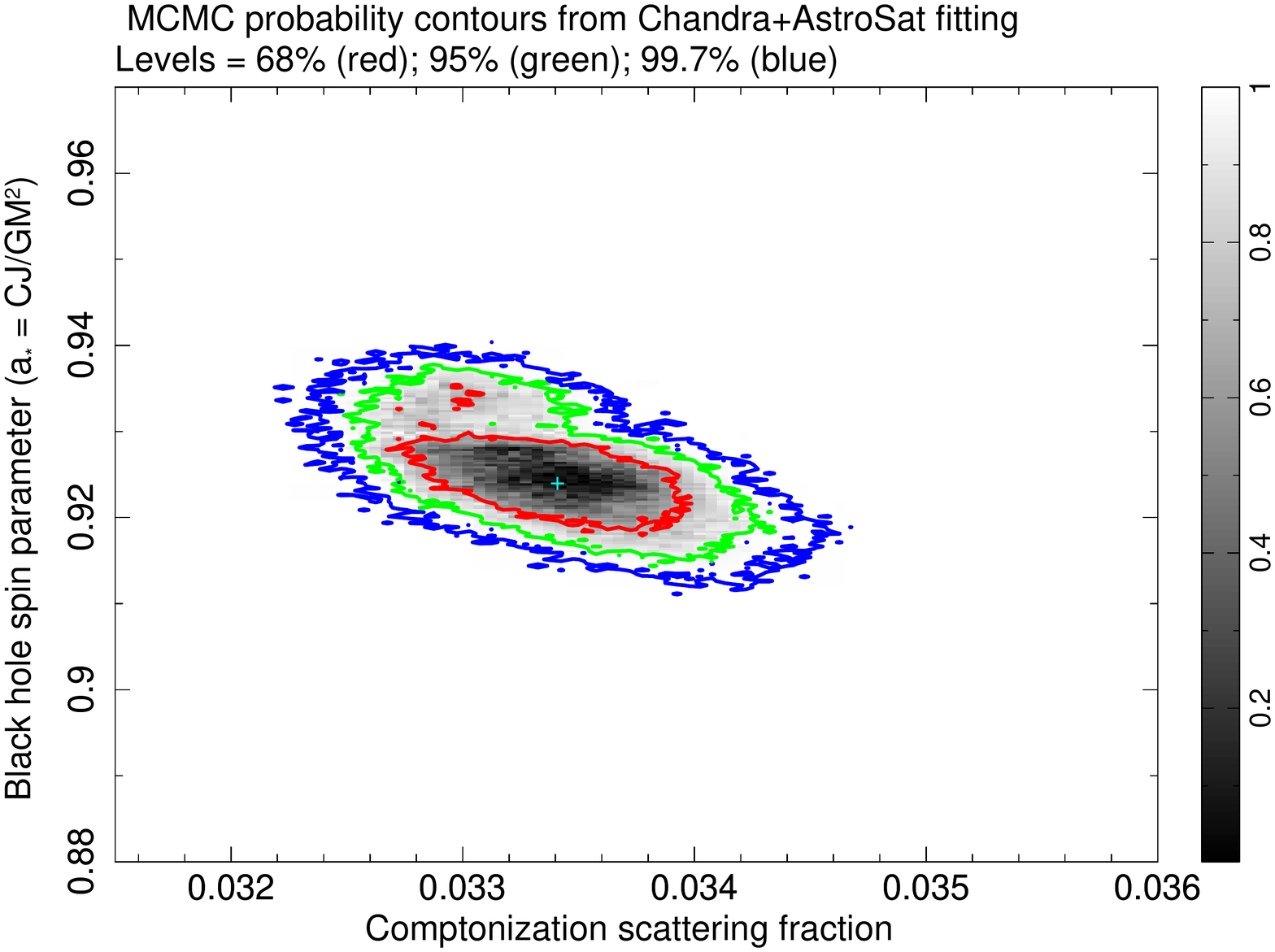}
\caption{Top left and top right panels show MCMC derived 1$\sigma$ (red), 2$\sigma$ (green) and 3$\sigma$ (blue) integrated contours from marginal probability distribution of black hole spin parameter as a function of spectral hardening factor obtained from \chan{}/HEG spectral fitting as well \chan{}/HEG+\asat{}/SXT+\asat{}/LAXPC joint spectral fitting respectively. Bottom left and bottom right panels show (for \chan{}+\asat{}) the black hole spin parameter as a function of the {\tt simpl} powerlaw index and the {\tt simpl} Comptonization scattering fraction. }  
\label{cont}
\end{center}
\end{figure*}

To check the relation between the spin parameter and the spectral hardening factor, we extract MCMC-derived 1$\sigma$, 2$\sigma$ and 3$\sigma$ integrated contours of the black hole spin parameter as a function of the spectral hardening factor from marginal probability distribution of fitted spectral parameters from \chan{}/HEG and \chan{}/HEG+\asat{}/SXT+LAXPC, and plot them in the top left and top right panels of Figure \ref{cont} respectively. From both panels, it is clear that the spectral hardening factor is constrained in the range of $1.5-1.7$ with the 99.7\% confidence level. This is a little lower than the typically used value of the spectral hardening, i.e., 1.7, but it is in agreement with the fact that both \chan{} and \asat{} spectra are observed during high soft state. This is because \citet{sa13} showed that during the high soft state and at an X-ray flux $> 10^{-8}$ ergs s$^{-1}$ cm$^{-2}$ in the $0.1-10$ keV range, the spectral hardening factor is usually less than 1.7.

To check if there exists degeneracy among model components, e.g., degeneracy between {\tt simpl} and {\tt kerrbb} model parameters, in Figure \ref{cont} we plot 1$\sigma$ and 2$\sigma$ contours of the MCMC-derived black hole spin parameter from {\tt kerrbb} model versus (1) MCMC-derived {\tt simpl} photon powerlaw index (bottom left panel), and versus (2) {\tt simpl} Comptonization scattering fraction (bottom right panel), used to fit \chan{}/HEG+\asat{}/SXT+LAXPC spectra. Both plots show that the 3$\sigma$ range of photon powerlaw index between $2.7-3.0$ and Comptonization fraction between 3.2\%$-$3.5\%.

Such result implies that the black hole spin parameter obtained from our analysis is robust and with the measurement using independent instrument like \chan{}/HEG, \asat{}/LAXPC and \asat{}/SXT, it is well-constrained in the range of 0.88-0.96 with 3$\sigma$ limit.

\section{Discussion and conclusions}\label{Discussion}

In this work, we study the 2016 X-ray outburst from the black hole X-ray transient 4U 1630--47. Using \maxi{} and \swift{}/BAT one-day averaged lightcurves in different energy bands, we show that the outburst is visible in the $2-4$ keV and $4-10$ keV energy range and is marginally detected in the $15-50$ keV hard X-ray band only during the decay phase of the outburst. During the rising phase, the non-detection in  the hard X-ray as well in the radio band confirms that the source has undergone a soft X-ray outburst. 
Similar bursts were also seen previously from this source. Very close to the peak of the outburst, \asat{} and \chan{} observed the source with 94.6 ks and 30.9 ks exposure times, respectively. Although these observations were taken $\sim$20 days apart, using the HID and the CCD diagram from \maxi{} data, we show that the hard and soft color values at the times of \asat{} and \chan{} observations were not significantly different, and these two observations were performed during the canonical high soft spectral state of the source. To further confirm the spectral state, we model the \chan{}/HEG spectra, and find that the continuum is well fitted with a single relativistic, disk blackbody model. When we extend the broadband fitting up to 23 keV by including the \asat{}/LAXPC, we find that a very steep powerlaw (index $> 2.5$) is required at a marginal scattering fraction of 3-4\%. Using \chan{} grating spectra, we notice very strong absorption lines at $6.705_{-0.002}^{+0.002}$ keV and $6.974_{-0.003}^{+0.004}$ keV,
which we identify as the mildly blueshifted Fe XXV and Fe XXVI absorption lines corresponding to the rest frame energies at 6.697 keV and 6.966 keV respectively. If these blueshifted lines are originated from an outflowing wind, then the velocity of the wind would be 366 $\pm$ 56 km/s.
 
We find no evidence of a broad Fe emission line or any other signature of reflection features in either the \chan{} or \asat{} spectra. This is not surprising, as previously the reflection spectral component from 4U 1630--47 \citep{ki14}  was observed during the intermediate state, while our observations were performed in the more luminous high soft state. We may note that using \xmm{} spectra, \citet{di14} found an evidence of Fe emission line at the 2-10 keV unabsorbed flux of 2.12 $\times$ 10$^{-8}$ ergs ~s$^{-1}$ ~cm$^{-2}$. The 2-10 keV unabsorbed flux computed from our \chan{} and \asat{} joint spectral modelling is 1.33$^{+0.08}_{-0.05}$ ergs ~s$^{-1}$ ~cm$^{-2}$ which is lower by a factor of $\sim$2 than that reported by \citet{di14}. However, 2-10 keV unabsorbed flux during their Obs 3 (also see Table 2) matches with our value and rather than the Fe emission line, they found absorption lines during the Obs 3 similar to what we found in the current study. Therefore, the accretion geometry as observed from the \xmm{} and the present \chan{} grating spectra of 4U 1630--47 may be similar at similar flux level.    

Assuming a black hole mass of 10 $M_\odot$, the Eddington luminosity of 4U 1630--47 would be 1.26 $\times $ 10$^{39}$ ergs/s. The unabsorbed X-ray flux in $0.1-100$ keV inferred from the joint \chan{} and \asat{} spectral fitting is 2.81 $\times$ 10$^{38}$ ergs/s. Therefore, the source was accreting at $\sim 22$\% of the Eddington accretion rate during our observations. We further note that a very weak Comptonization component, inferred from our spectral fitting, implies a low hard X-ray flux available for reflection from the disk. Such a low fraction of hard X-ray flux may be a reason for the non-detection of a reflection spectral component.

There are two widely used techniques for the black hole spin measurement, based on (1) the modeling of the broad relativistic Fe emission line and other reflection features, and (2) the modeling of relativistically-modified thermal continuum spectra. In this paper, we use the latter one. For the thermal continuum modelling, the spectral data selection has the following critical criteria: (1) the continuum spectrum should have a dominant thermal disk-blackbody emission from a geometrically-thin and optically-thick accretion disk; (2) a significant presence of other spectral components, like the reflection or Comptonization continuum, is undesirable, since their presence complicates the spectral modelling, and introduces systematic uncertainties in spin measurement; and (3) the X-ray luminosity should be fairly high, but $\lsim 30$\% of the Eddington luminosity, so that accretion disk is not radiation pressure dominated, and the geometrically thin disk assumption still holds true. 
Our continuum spectrum, as observed with both \chan{} and \asat{}, satisfies all these criteria. This is because no reflection features are observed, X-ray luminosity is $\sim 22$\% of the Eddington luminosity, and $\sim 95$\% of the X-ray emission is from a geometrically thin disk. Therefore, our continuum spectra are ideal to determine the black hole spin.

\subsection{On black hole spin measurement}\label{spin}

By independently modeling \chan{}/HEG, \asat{} and \chan{}+\asat{} broadband spectra, and using MCMC simulations on fitted spectral parameters, we find that the 3$\sigma$ range of the black hole spin parameter is $0.88-0.96$, which indicates the presence of a rapidly-spinning black hole in 4U 1630--47. This is close to the spin parameter value of 0.985$^{+0.005}_{-0.014}$, estimated from modeling reflection spectra of 4U 1630--47 \citep{ki14}. Note that we fix $i$, $D$ and $M$ in our spectral fitting. Our assumption of $i = 64^\circ$ should be reasonable, because independent methods have argued or predicted similar values (e.g., $i \lsim 70^\circ$, see Section~\ref{Introduction}; $i = 64^\circ\pm2^\circ$; \citet{ki14}). As argued in Section~\ref{Introduction}, our assumption of $D = 10$~kpc should also be reasonable. While our assumption of $M \approx 10 M_\odot$ was adopted from the estimated mass reported in \citet{se14}, the absence of a dynamical measurement of mass implies the lack of a confirmed value. We, therefore, considering reasonable limits of the typical stellar black hole mass range, fix the black hole mass at $5 M_\odot$ and $15 M_\odot$ in our joint spectral modelling of \chan{} and \asat{}, fix the best-fit spin values and keep the distance free to vary between 2 kpc and 50 kpc. The best-fit returns an acceptable $\chi^2$/dof = 533/477 (1.12) and 549/477 (1.15) for the black hole mass of $5 M_\odot$ and $15 M_\odot$ respectively. This implies for a reasonable range of black hole masses and distances, the spectral fitting supports the high black hole spin. Therefore, our finding of a rapidly-spinning black hole in 4U 1630--47 should be reliable.

\acknowledgments

We thank the referee for constructive comments which improve the quality of the manuscript. We acknowledge the strong support from Indian Space Research Organization (ISRO) during the instrument building, testing, software development and mission operation. We also acknowledge the support from the LAXPC Payload Operation Center (POC), TIFR, Mumbai. This work has used the data from the Soft X-ray Telescope (SXT) developed at TIFR, Mumbai, and the SXT POC at TIFR is thanked for verifying and releasing the data via the ISSDC data archive and providing the necessary software tools. The scientific results reported in this article are based on observations made by the Chandra X-ray Observatory. This research has made use of the software provided by the Chandra X-ray Center (CXC) in the application packages CIAO, ChIPS, and Sherpa. This research has made use of the MAXI data provided by RIKEN, JAXA and the MAXI team. MP thanks TIFR, Mumbai for the support and hospitality during this work, and acknowledges Royal Society-SERB Newton International Fellowship support funded jointly by the Royal Society, UK and the Science and Engineering Board of India (SERB) through Newton-Bhabha Fund. MP and PG acknowledges the support from the UGC-UKIERI Phase 3 thematic partnership grant 2017/18 at the School of Physics and Astronomy, University of Southampton, UK. PG thanks for the support and funding from STFC (ST/R000506/1).

\end{document}